\documentclass[letterpaper]{aa}  
\usepackage{graphicx}
\usepackage{natbib}

\begin{document}
   \title{NLTE model calculations for the solar atmosphere with an iterative treatment of opacity distribution functions}

   \author{M. Haberreiter
         \inst{1}\thanks{Present affiliation: LASP, University of Colorado, 1234 Innovation Drive, Boulder, CO, 80303, USA}
          \and
          W. Schmutz
         \inst{1}
          \and
          I. Hubeny
         \inst{2}}

   \offprints{haberreiter@lasp.colorado.edu}

   \institute{Physikalisch-Meteorologisches Observatorium Davos, World Radiation Center,
              Dorfstrasse 33, CH-7260 Davos Dorf\\
              \email{margit.haberreiter@pmodwrc.ch, werner.schmutz@pmodwrc.ch}
         \and
             Steward Observatory, University of Arizona, Tucson, AZ 85721, USA\\
             \email{hubeny@as.arizona.edu}
             }

   \date{}

  \abstract
   {Modeling the variability of the solar spectral irradiance is a key factor for understanding the solar
influence on the climate of the Earth.}
   {As a first step to calculating the solar spectral irradiance variations we reproduce the solar spectrum for the quiet Sun over a broad wavelength range with an emphasis on the UV.}
   {We introduce the radiative transfer code COSI which calculates solar synthetic spectra under conditions of non-local thermodynamic equilibrium (NLTE). { A self-consistent simultaneous solution of the radiative transfer and the statistical equation for the level populations guarantees that the correct physics is considered for wavelength regions where the assumption of local thermodynamic equilibrium (LTE) breaks down.} The new concept of iterated opacity distribution functions (NLTE-ODFs), through which all line opacities are included in the NLTE radiative transfer calculation, is presented.}
   {We show that it is essential to include the line opacities in the radiative transfer to reproduce the solar spectrum in the UV.}
   {Through the implemented scheme of NLTE-ODFs the COSI code is successful in reproducing the spectral energy distribution of the quiet Sun.}
   \keywords{Radiative transfer -- Sun: atmosphere -- Sun: chromosphere -- Sun: UV radiation -- Line: profiles -- Atomic data}
   \titlerunning{NLTE calculations of the solar atmosphere}
   \authorrunning{Haberreiter et al.}
   \maketitle

\section{Introduction}
{While the total solar irradiance can not be the cause of the climate change over the past 20 years \citep{Lockwood2007}}, it is known to influence the Earth''s pre-industrial climate \citep{Labitzke2005,KilHaigh2005,Shindell2003}. However, it has not yet been fully identified how exactly the irradiance variations affect the climate. It has been argued that direct forcing through the variation of the total solar irradiance is too small to produce the observed climate changes as, e.g., observed during the little ice age \citep{Shindell2001}. One of the prime candidates for an indirect effect is the variation of the UV part of the solar spectrum, which varies much more than the total irradiance. In particular it has already been shown that H\,{\sc{i}} 121.6\,nm (Lyman-$\alpha$) and the Herzberg band around 200\,nm affect ozone and the temperature in the stratosphere \citep{Rozanov2006, Rozanov2002, Egorova2004}. 

To understand the solar spectrum variations it is essential to consider all the relevant physical processes in the formation of the solar spectrum. This research has already been carried out by several authors who use different approaches and focus on various wavelengths ranges. Calculations of solar spectra in 1D over broad wavelength ranges were carried out by \citet{Kurucz1991,Kurucz2005e} based on LTE radiative transfer calculations with the ATLAS9 and ATLAS12 code in plane-parallel symmetry. Furthermore, \citet{Vernazza1981}, \citet{Allende2003b,Allende2003a}, \cite{Fontenla1999,Fontenla2007}, and \cite{AvrettLoeser2008ApJS} calculated solar spectra in NLTE and plane-parallel symmetry for the visible and IR. Improved understanding of the solar spectrum formation has been achieved by the detailed study of individual spectral lines by \cite{Ayres2006} and \cite{Collet2005} who calculated models out of local thermodynamic equilibrium (NLTE). Furthermore,
\citet{HubenyLites1995}, \citet{Uiten2001,Uiten2002}, \cite{Fox2004}, and \cite{Fontenla1999,Fontenla2007} improved the synthetic calculations by including partial redistribution (PRD). Finally \citet{ShortHaus2005} computed model spectra of cool stars in NLTE and spherical symmetry. Furthermore, 3D NLTE radiative transfer calculations based on 3D hydrodynamic simulations have been carried out, e.g. by \cite{Asplund2000}, \cite{Koesterke2008}. Here we limit ourselves to radiative transfer calculations in 1D, as our reconstruction approach of the spectral solar irradiance is based on 1D model atmosphere structures of solar surface features and their distribution on the solar surface \citep{Wenzler2006,Wenzler2005,Hab2005AdSpR,Krivova2005}.

The formation of the solar spectrum is, in some spectral ranges, dominated by an immense number of spectral lines, e.g. the iron like elements in the UV, also known as UV {\em line haze}. Due to the characteristics of the solar temperature structure there are two effects of the line opacity, which depend on the height of line formation, that have to be distinguished. One effect is the line blocking in the photosphere leading to a decrease of intensity. The other effect is the excess of intensity due to emission lines in the chromosphere. Because of NLTE effects, i.e. the illumination from above, the chromospheric emission lines can lead to overionization at lower layers. In this paper we do not focus on the chromospheric emission lines apart from Lyman-$\alpha$, as they can only be correctly calculated if all atomic processes are treated in full NLTE, but discuss the line blocking in the photosphere. 

As shown by \cite{Collet2005}, the lack of photospheric line blocking leads to an incorrect excess of ionization. The authors find differences between calculations including and excluding line opacity in NLTE calculations with a modified version of the statistical equilibrium code MULTI \citep{Carlsson1986} for cool stars with solar metallicity. They show that the inclusion of sampled background line opacities decreases the mean intensity field at a given depth and thereby reduces the radiative ionization rates. The authors find that the decrease of ionization in turn changes the line strength and leads to a Fe abundance that is 0.1-0.15 dex higher than in calculations without line blocking. From this it follows that all opacities contributing to the line blocking have to be accounted for in NLTE radiative transfer calculations.

There have already been a number of approaches for including line opacities in the background continuum opacity. \cite{AndersonAthay1989} and \cite{Anderson1989} first introduced the concept of super levels that allows for numerous levels of certain atoms with similar energies, which are assumed to be in LTE with respect to each other. This approach was later adopted by \cite{DreizlerWerner1993} and \cite{HubenyLanz1995}, who used ODFs for the transitions between the superlevels. Describing line opacities with a Monte-Carlo evaluation has been applied by \cite{SchmuLeiHubVog1991}, \cite{SchaererSchmutz1994}, \cite{Schmutz1997}, and \cite{DeMarcoSchmutz1999} in the calculation of the mass loss of hot stars. Furthermore, \cite{ShortHaus2005} include more than 100,000 spectral lines or iron-like ions that blanket the UV-band self-consistently in the NLTE calculation. \citet{Collet2005} and \cite{AvrettLoeser2008ApJS} include the line opacities in the background continuum opacity via sampled line opacities. Our approach improves the self-consistency of the NLTE calculation by including the line opacities from all known transitions of all ions into the radiative transfer calculations by means of iterated opacity distribution functions which include some NLTE effects, and therefore, we name these functions NLTE-ODFs. 

The main purpose of this paper is not so much a detailed analysis of line profiles, but the calculation of realistic solar energy distribution over a broad wavelength range with an emphasis on the UV and to be able to reproduce the solar observations and reference spectra over a broad wavelength range.

In the next section we describe the radiative transfer code COSI, the model atoms, the model atmosphere, the implemented cross sections and the different sets of solar abundances used in our calculations. Then, in Sect.\,\ref{sec:lte} we validate the performance of COSI in LTE against the ATLAS12 calculations. In Sect.\,\ref{sec:nlte} we describe the new concept of NLTE-ODFs and present the NLTE calculations carried out with COSI. In Sect.\,\ref{sec:disc} the results are discussed and finally in Sect.\,\ref{sec:concl} the conclusions are presented.
\section{COSI}
\begin{figure}[!t]
\centering
\includegraphics[width=\linewidth]{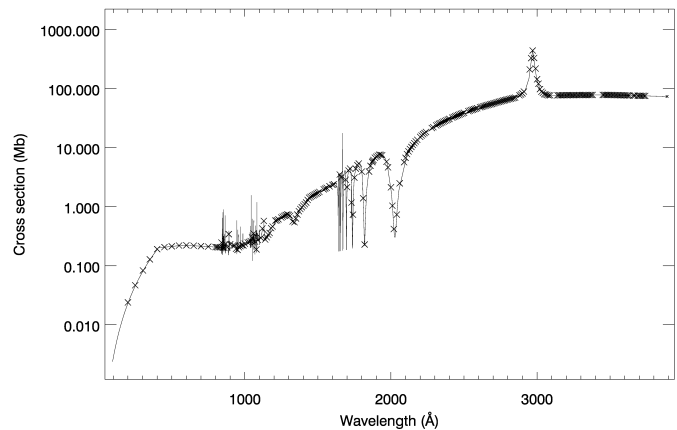}
\caption{Photoionization cross section for the third level of neutral Mg ({\em{MgI 3}} in Table\,\ref{tab:levels}) implemented in COSI. The solid line is the original data from the Opacity Project and the crosses are the data interpolated to the frequency grid applied in COSI. \label{fig:cs}}  
\end{figure}
\subsection{Introduction}
COSI (COde for Solar Irradiance) is a combination of two codes, a model atmosphere code, developed by \cite{HamannSchmutz1987} and \cite{SchmuHamWess1989}, hereafter COSIMA, and the spectrum synthesis program SYNSPEC, going back to \cite{Hubeny1981} and further developed by \cite{Hubeny1988} and \cite{HubenyLanz1995}. The new method introduced by this paper is the use of NLTE-ODFs for including the effects of line blanketing replacing the Monte-Carlo evaluation of the line opacity that was used in earlier applications \citep{SchmuLeiHubVog1991,SchaererSchmutz1994,Schmutz1997,DeMarcoSchmutz1999}. COSIMA calculates the NLTE populations for a set of specified atomic levels by solving the radiative transport equations simultaneously with the equations for statistical equilibrium. The radiative transfer is solved in spherical symmetry \citep{Mihalas1978,Peraiah2001}, which yields a more reliable emerging intensity at the limb than a plane-parallel geometry and allows to calculate line of sights at and beyond the solar limb \citep{Haberreiter2008}, the contribution of which becomes increasingly important for UV/EUV wavelengths which are formed higher in the solar atmosphere.

The spectral synthesis code takes the NLTE level populations for the explicit levels from COSIMA and assumes LTE populations for the non-explicit atomic levels. The latter are calculated in LTE relative to the NLTE ground states. Population numbers of lines that connect a lower explicit NLTE level and a non-explicit upper level are assumed to be in LTE with respect to the explicit level. Partition functions are used for the calculation of the ionization equilibria in LTE, following the Saha-Equation. The partition functions for the elements H to Zn they are taken from \cite{Traving1966} and for the heavier elements the coefficients for polynomial fits are taken from \cite{Irwin1981}. This allows us to considers all bound-bound transitions from hydrogen to thallium, leading to more than $2.8 \times 10^6$ lines. The linelist was provided by \citet{Kurucz2006}.
\subsection{Atmosphere structures}
For the validation of COSI in LTE we apply the atmosphere structure for the quiet Sun by \citet[hereafter ASUN]{Kurucz1991}, which is calculated in LTE and has a monotonic outward decreasing temperature profile with the outer boundary at the temperature minimum. 

The NLTE spectra are calculated with the semi-em\-pirical structure of the solar atmosphere for average supergranule cell interior (Model C) by \citet{Fontenla1999}, further referred to as F1999.
\subsection{Model atoms}
\begin{figure*}[!tthhh]
\centering
\includegraphics[width=.49\linewidth]{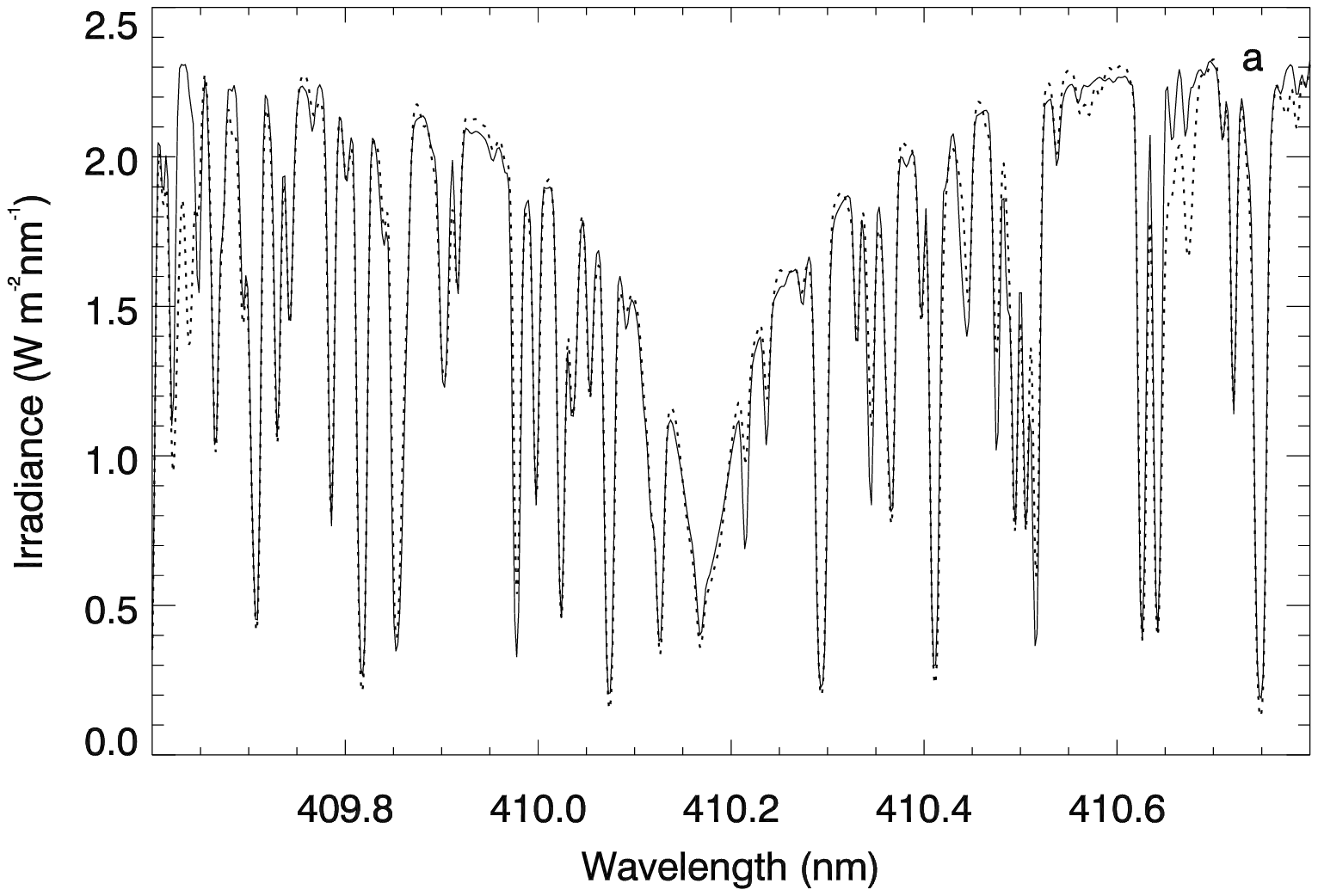}
\includegraphics[width=.49\linewidth]{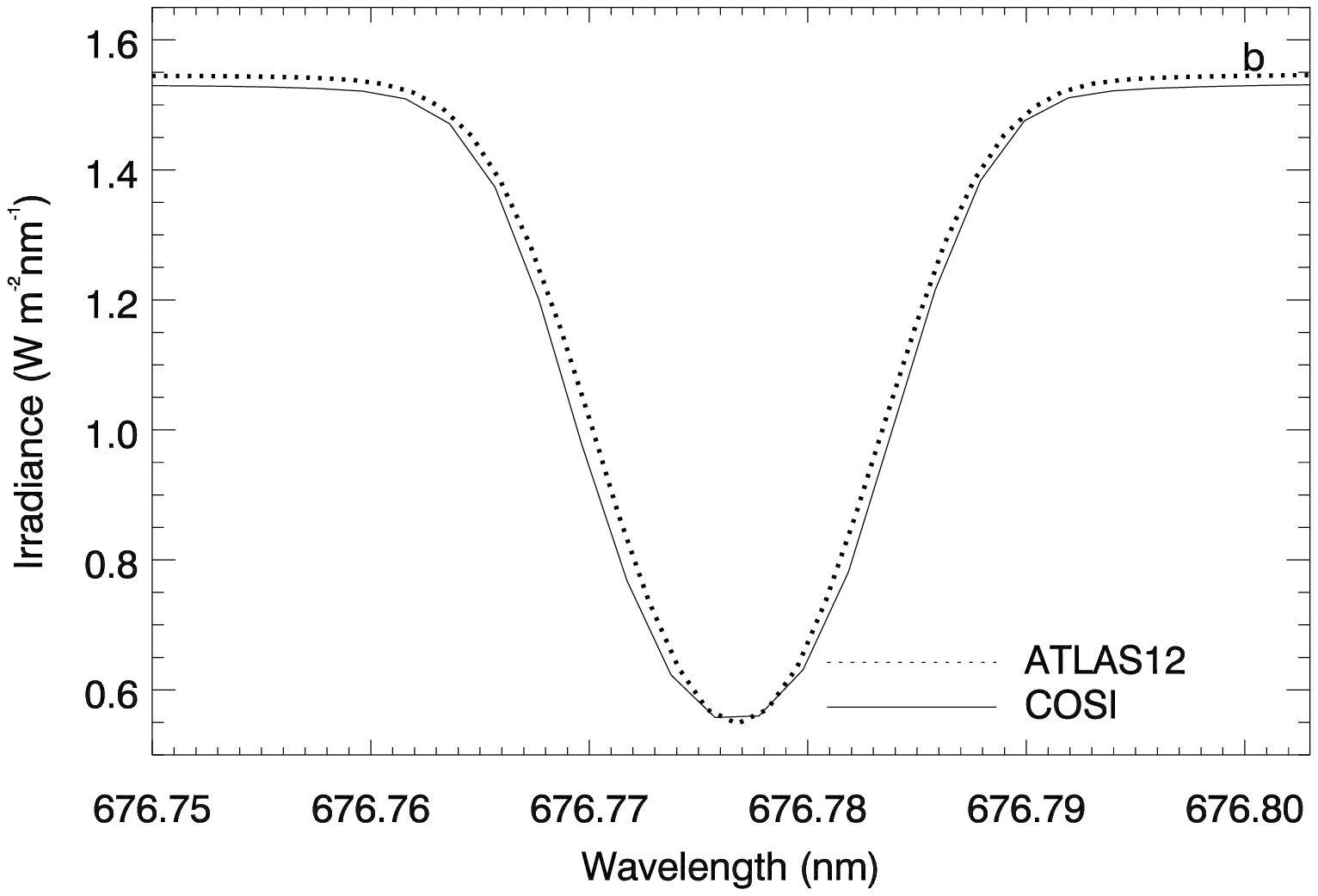}
\caption{Comparison of ATLAS12 calculation by \citet{Kurucz2005e} (dotted line) and COSI calculation (solid line) for {H\,{\sc{i}}\,\,410.1\,nm} (panel a) and {Ni\,{\sc{i}}\,\,676.8\,nm} (panel b), using the same abundances, atmosphere structure, microturbulence (1.5 km/s), Gaussian macroturbulence (FWHM=1.5 km/s) and rotation broadening (2 km/s).
 \label{fig:LTE_lines}}  
\end{figure*}
The atomic model of hydrogen consists of negative hydrogen, 10 levels of neutral hydrogen each representing the quantum numbers up to $n=10$, and the ion. Helium is modeled by 10 levels of neutral helium and one level for the ground state of singly ionized helium. Furthermore, we have implemented all metals up to zinc in the NLTE calculation. Currently we account for a total number 61 ions and 114 explicit NLTE le\-vels as given in Table\,\ref{tab:levels}. The metal levels are generally selected according to their importance for the solar UV continuum opa\-city, e.g. their ionization edges are in the wavelength range from 100 to 400\,nm. 

All bound-bound and bound-free transitions connecting the explicit levels are fully included in the NLTE calculation. In the present implementation the millions of additional transitions are set to LTE relative to the ground level of the corresponding ionization stage, which is allowed to deviate from LTE and explicitly calculated. Thus the LTE transitions are to some extent affected by NLTE conditions and in turn determine the NLTE-ODFs.  

For line transitions connecting an explicit lower level with a non-explicit upper level the latter is calculated in LTE relative to the NLTE population of the lower level. Here however, an inaccuracy occurs for the chromospheric lines that are calculated in LTE, because this leads to an overestimation of the line emissivity. To avoid this excess of emission in the spectral synthesis the line emissivity for each depth point outward of the temperature minimum is calculated with the excitation temperature set to the value of the temperature minimum, i.e.\  
as $\epsilon_{L,\lambda}=B_{\lambda, T_{\mathrm{min}}}\kappa_{L,\lambda}$, with $B$ being the Planck function and $L$ the depth point index of the atmosphere structure. We are aware of the fact that this approach clearly underestimates the chromospheric emission lines. However, the large excess of the emission lines due to their wrong LTE calculation is not the reasonable solution either. This shortcoming will be addressed in the future by including all the relevant levels in the NLTE calculation.
\subsection{Cross sections}
\subsubsection{Radiative processes}
The bound-free and free-free radiative processes involving negative hydrogen are implemented according to \citet{John1988}. The photoionization cross section for the ground state of hydrogen is calculated according to \citet{Mihalas1967} and the cross sections of the other hydrogen levels are adopted from \citet{Seaton1960}.
The bound-free radiative processes for helium are calculated according to \cite{Koester1985}. Free-free cross sections of positively charged ions and electrons are calculated with an hydrogenic approximation using the cross sections by \cite{Berger1956} and \cite{KarzasLatter1961}.

Table\,\ref{tab:levels} gives the sources for the photoionization cross sections for all the 114 explicit NLTE-levels accounted for in COSI. For the ions C\,{\sc{i}}, Na\,{\sc{i}}, Mg\,{\sc{i}}, Al\,{\sc{i}}, Si\,{\sc{i}}, S\,{\sc{i}} and Ca\,{\sc{i}} the photoionization cross sections are taken from the Opacity Project \citep{Seaton1994} and the ones for Fe\,{\sc{i}} are provided by \cite{Bautista1997,BautistaPradhan1997}. We linearly interpolate the complex photoionization cross sections to the slightly coarser frequency grid employed in COSI \citep{Haberreiter2003}. As an example, the cross section of Mg\,{\sc{i}} is given in Fig.\ref{fig:cs}. As theo\-re\-tical ionization energies are slightly off from the observed ones, we corrected the theoretical values to agree with the measured energies taken from the National Institute of Standard and Technology (NIST) Atomic
Spectra Database. Also, Thomson scattering and Rayleigh scattering are included in COSI. 
\subsubsection{Collisional processes}
In COSI the collisional processes involving negative hydrogen include the collisional detachment by electrons and neutral hydrogen and charge neutralization with protons as given by \citet{LambertPagel1968}, Eq. (2b,c,e) therein. The collisional bound-bound cross sections of neutral hydrogen, the optically permitted transitions of helium, as well as of the neutral and ionized metals are calculated using the formula of
\citet{Jeff1968}, p.118, Eq. 6.24. Collisional processes of neutral hydrogen other than with negative hydrogen are not yet included.
\subsection{Solar abundances}
\begin{figure}[!hhht]
\centering
\includegraphics[width=\linewidth]{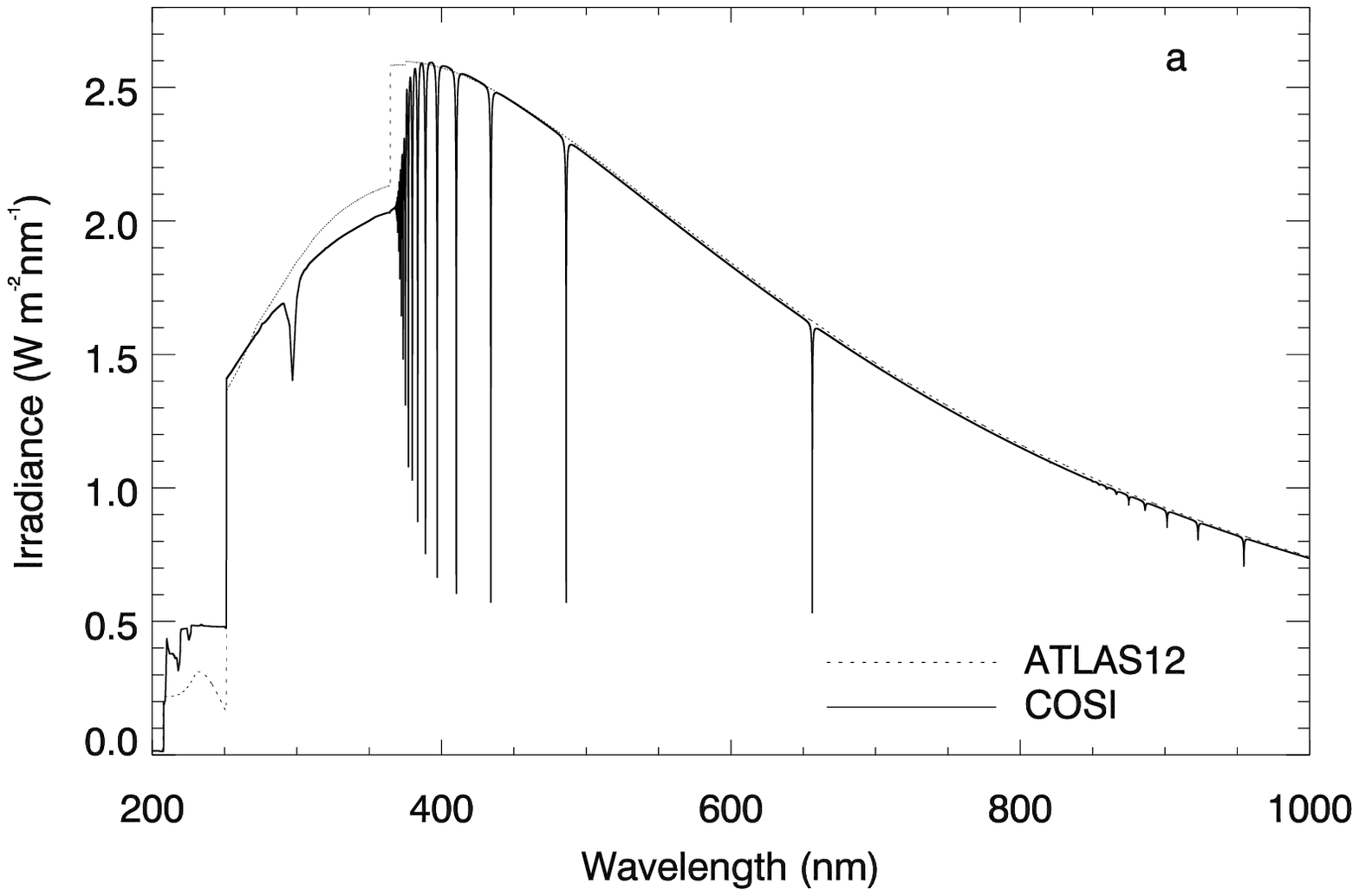}
\includegraphics[width=\linewidth]{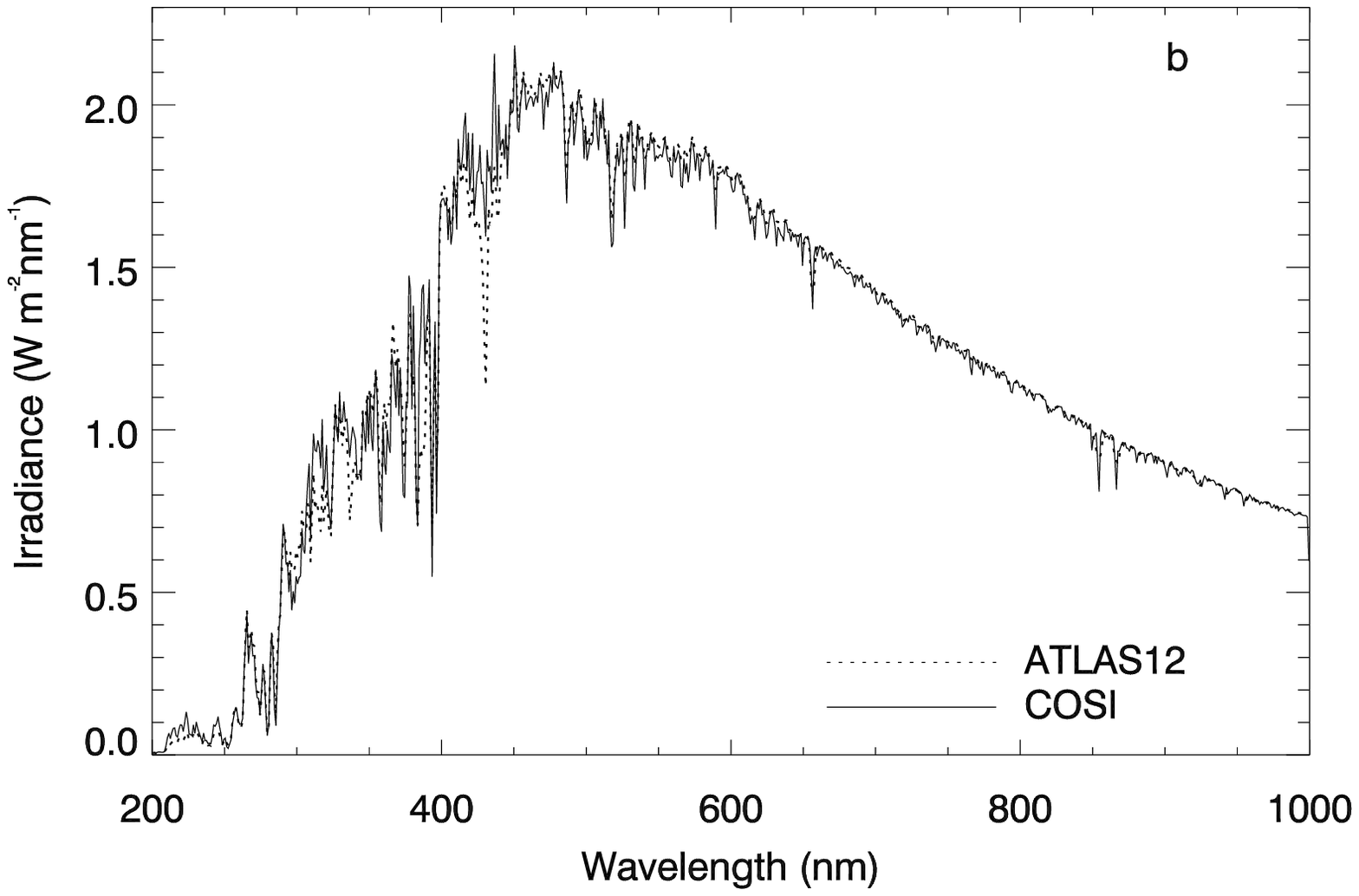}
\includegraphics[width=\linewidth]{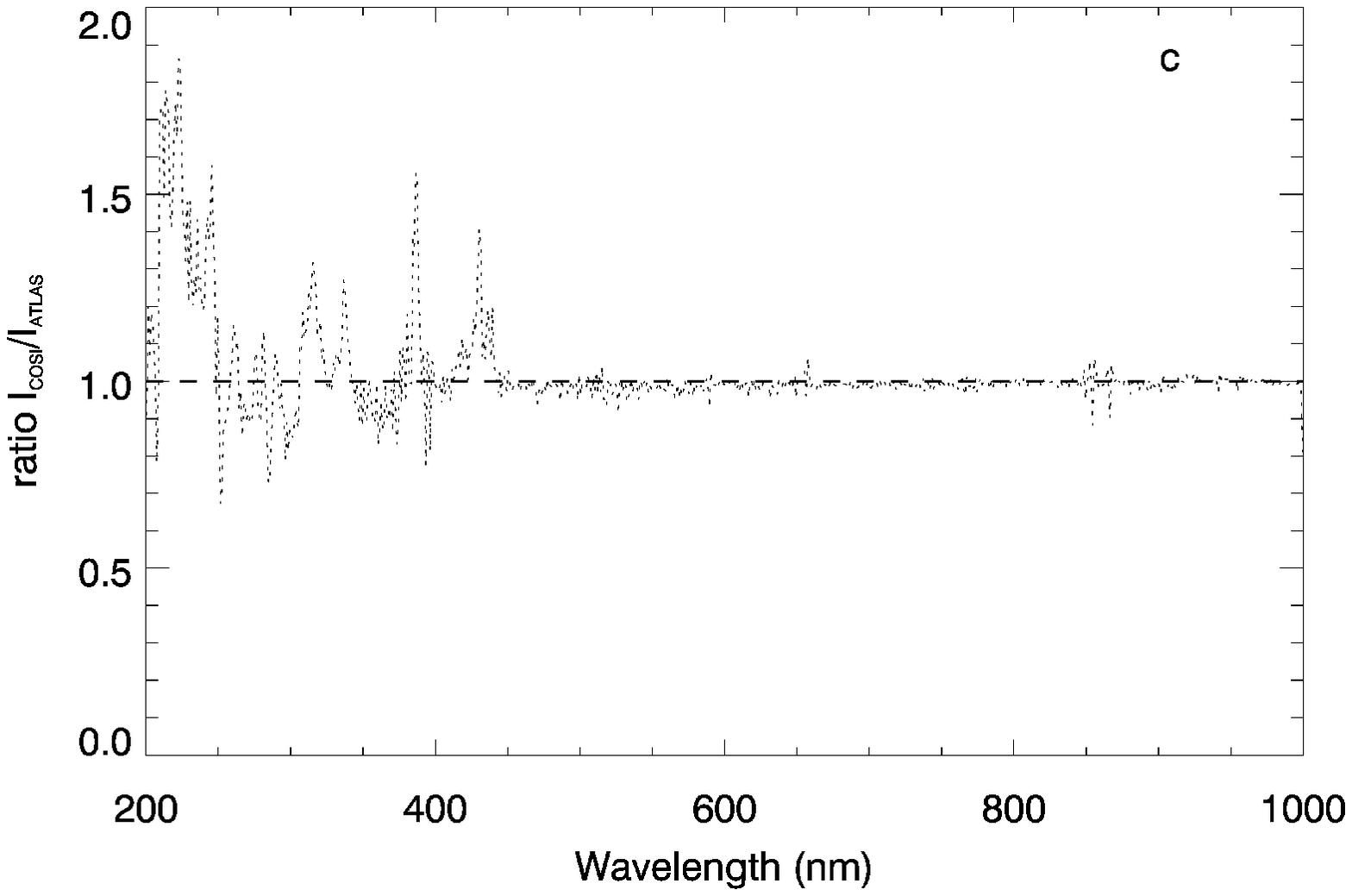}
\caption[Spectrum in LTE]{Panel (a) shows the comparison of the continuum calculated with the ATLAS12 code (dotted line) and continuum plus hydrogen lines calculated with COSI in LTE (solid line), based on identical atmosphere structure and abundances. The difference shortward of the Balmer jump at 364.5\,nm is mainly due to the photoionization cross section of the third Mg\,{\sc{i}} level, shown in Fig.\,\ref{fig:cs}, apparently not accounted for in the ATLAS12 calculation. Panel (b) gives the corresponding synthetic LTE spectra averaged with a 1-nm boxcar. Panel (c) shows the ratio between the COSI and the ATLAS12 calculation. \label{fig:LTEIP0}}  
\end{figure}
\begin{table}
\begin{center}
\caption{List of applied solar abundances relative to H by number for the elements (EL) with atomic number A and atomic mass m$_\mathrm{A}$ \label{tab:abndc}} 
{\small{
    \begin{tabular*}{.47\textwidth}{lrr@{.}l
    r@{.}l@{\,\,E\,}l
    r@{.}l@{\,\,E\,}l
    r@{.}l}
      \hline
      \noalign{\smallskip}
      {{El}}  & 
      {{A}} &
      \multicolumn{2}{c}{m$_\mathrm{A}$}  & 
		  \multicolumn{3}{c}{$\mathrm{K91}$} &
      \multicolumn{3}{c}{$\mathrm{F2007}$} &   
      \multicolumn{2}{c}{$\frac{\mathrm{F1999}}{\mathrm{K91}}$} \\      
      
      \noalign{\smallskip}
      \hline
      \noalign{\smallskip}                           
      H    &1& 1&008  &1&000&0      & 1&000 & 0  &1&000\\
      He   &2& 4&003  &9&769&-2     & 1&000 &-1  &1&024\\
      Li   &3& 6&941  &1&447&-11    & 1&122 &-11 &0&775\\
      Be   &4& 9&012  &1&414&-11    & 2&399 &-11 &1&696\\
      B    &5& 10&810 &3&985&-10    & 5&012 &-10 &1&258\\
      C    &6& 12&011 &3&635&-4     & 2&455 &-4  &0&675\\
      N    &7& 14&007 &1&123&-4     & 6&026 &-5  &0&536\\
      O    &8& 16&000 &8&521&-4     & 4&571 &-4  &0&536\\
      F    &9& 18&918 &3&634&-8     & 3&631 &-8  &0&999\\
      Ne  &10& 20&179 &1&231&-4     & 6&981 &-5  &0&562\\
      Na  &11& 22&990 &2&140&-6     & 1&479 &-6  &0&691\\
      Mg  &12& 24&305 &3&806&-5     & 3&388 &-5  &0&890\\
      Al  &13& 26&982 &2&954&-6     & 2&344 &-6  &0&793\\
      Si  &14& 28&086 &3&552&-5     & 3&236 &-5  &0&911\\
      P   &15& 30&974 &2&822&-7     & 2&291 &-7  &0&812\\
      S   &16& 32&060 &1&624&-5     & 6&918 &-6  &0&426\\
      Cl  &17& 35&453 &3&166&-7     & 3&162 &-7  &0&999\\
      Ar  &18& 39&948 &3&635&-6     & 1&514 &-6  &0&417\\
      K   &19& 39&098 &1&320&-7     & 1&202 &-7  &0&911\\
      Ca  &20& 40&080 &2&293&-6     & 2&042 &-6  &0&890\\
      Sc  &21& 44&956 &1&260&-9     & 1&122 &-9  &0&890\\
      Ti  &22& 47&900 &9&783&-8     & 7&943 &-8  &0&812\\
      V   &23& 50&941 &1&001&-8     & 1&000 &-8  &0&999\\
      Cr  &24& 51&996 &4&683&-7     & 4&365 &-7  &0&932\\
      Mn  &25& 54&938 &2&457&-7     & 2&455 &-7  &0&999\\
      Fe  &26& 55&847 &4&683&-5     & 2&818 &-5  &0&602\\
      Co  &27& 58&933 &8&327&-8     & 8&318 &-8  &0&999\\ 
      Ni  &28& 58&700 &1&780&-6     & 1&698 &-6  &0&954\\
      Cu  &29& 63&546 &1&624&-8     & 1&622 &-8  &0&999\\
      Zn  &30& 65&380 &3&985&-8     & 3&304 &-8  &0&999\\
      \noalign{\smallskip}
      \hline
      \noalign{\smallskip}
    \end{tabular*}
}}
\end{center}
\end{table}
The study of solar abundances is an ongoing research topic and various data sets are used by different authors. For the synthetic spectra presented in this paper, the abundances used by \citet{Kurucz1991} (K91) were applied in the LTE COSI calculations using the atmosphere structure ASUN. The abundances used by \citet{Fontenla2007} (F2007), which are based on \cite{Asplund2005b}, are employed for the calculations using the F1999 atmosphere structure. Table\,\ref{tab:abndc} lists both abundance sets and their ratio for the elements implemented in the atomic model. K91 is used in the COSI LTE calculations. F2007 is the abundance used in all the NLTE calculations presented in this paper.

\section{Code validation in LTE}\label{sec:lte}
In order to validate the overall performance of COSI in LTE we calculated the emergent flux of the quiet Sun with COSI, applying the same atmosphere structure (ASUN), abundances (K91), and values for microturbulence (1.5\,km/s), macroturbulence (Gaussian FWHM=1.5\,km/s) and rotation broadening (2\,km/s) as \citet{Kurucz2005e} for the ATLAS12 calculations. The rotation broadening has been implemented as given by \cite{Gray1992}. Fig.\,\ref{fig:LTE_lines} shows the calculations with ATLAS12 (dotted line) and COSI (solid line) for the spectrum centered at {H\,{\sc{i}}\,\,410.1\,nm} (panel a) and {Ni\,{\sc{i}}\,$\lambda$\,676.8\,nm} (panel b). 
The line profiles of {H\,{\sc{i}}\,\,410.1\,nm} and in particular the wings of this line are almost identical, showing that in COSI the radiative transfer and the Stark broadening are computed consistently with the ATLAS12 code. The absolute flux level of the continuum at the {Ni\,{\sc{i}}} wavelength is slightly less than in the ATLAS12 calculation. As a result the line profile calculated with COSI appears slightly broader than the ATLAS12 calculation. However, the profiles are practically identical when related to the same continuum level.

The same calculation over the wavelength range from 200 to 1000\,nm is shown in Fig.\,\ref{fig:LTEIP0}. In panel (a) the continuum irradiance spectrum calculated with ATLAS12 (dotted line) and with COSI (solid line) are shown, where in the latter the hydrogen lines are included. Panel (b) shows the corresponding synthetic spectra including all spectral lines. To allow a better comparison the high resolution line spectra have been convolved with a 1-nm boxcar filter. Generally, the calculations show a good agreement, and the overall shape of the continuum spectrum is well reproduced. The strongest discrepancies are between $\sim$350 and 430\,nm. The differences between the spectra are due to missing molecular lines in COSI and due to different photoionization cross sections as evident from panel (a). In Fig.\,\ref{fig:LTEIP0} panel (c), the ratio between the COSI and ATLAS12 line spectra is given for each 1-nm bin. Below 420\,nm the ratio indicates some deviation while for longer wavelengths the ratio is very close to unity. 

\begin{figure*}[!t]
\centering
\includegraphics[width=.49\linewidth]{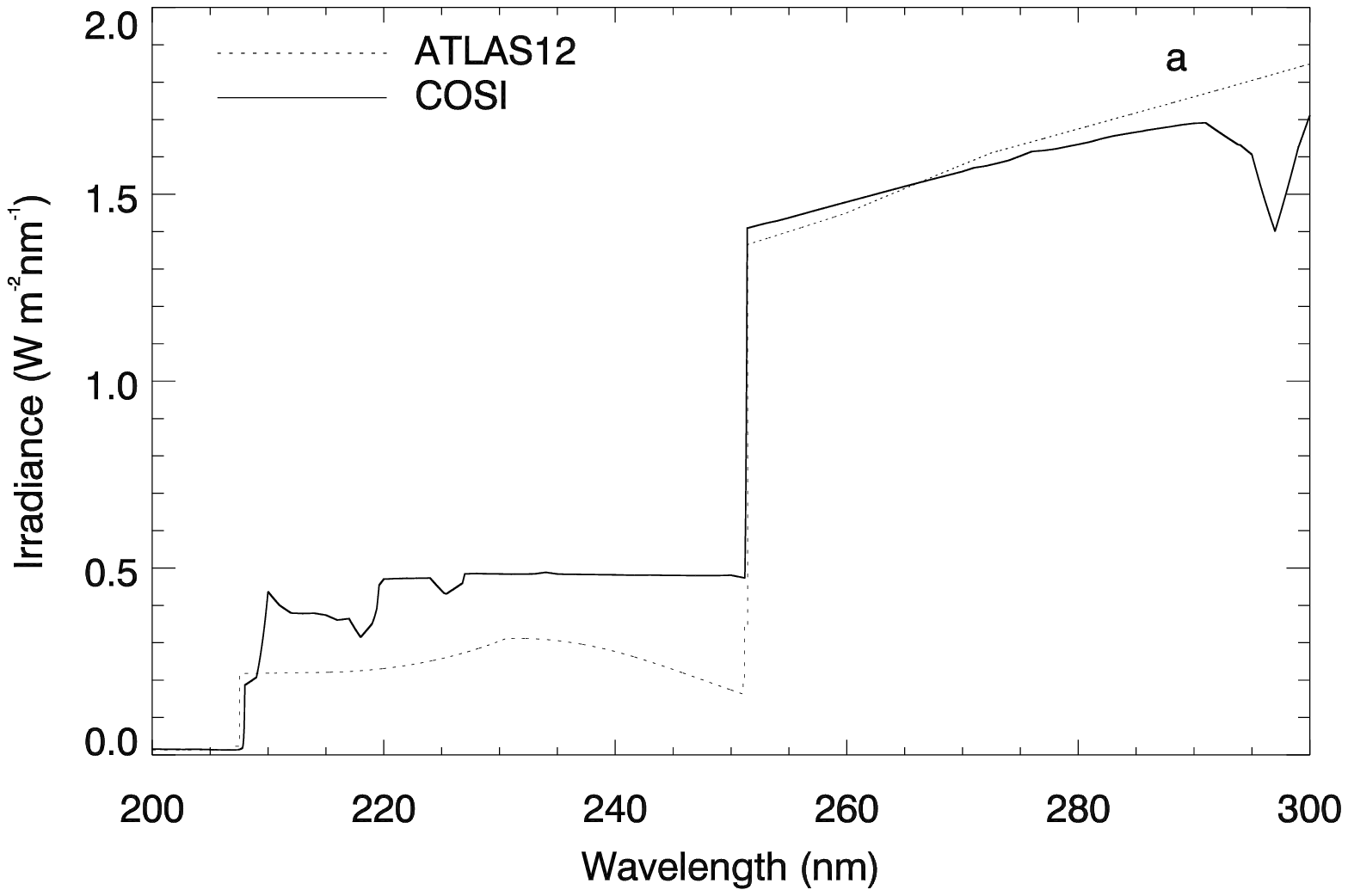}
\includegraphics[width=.49\linewidth]{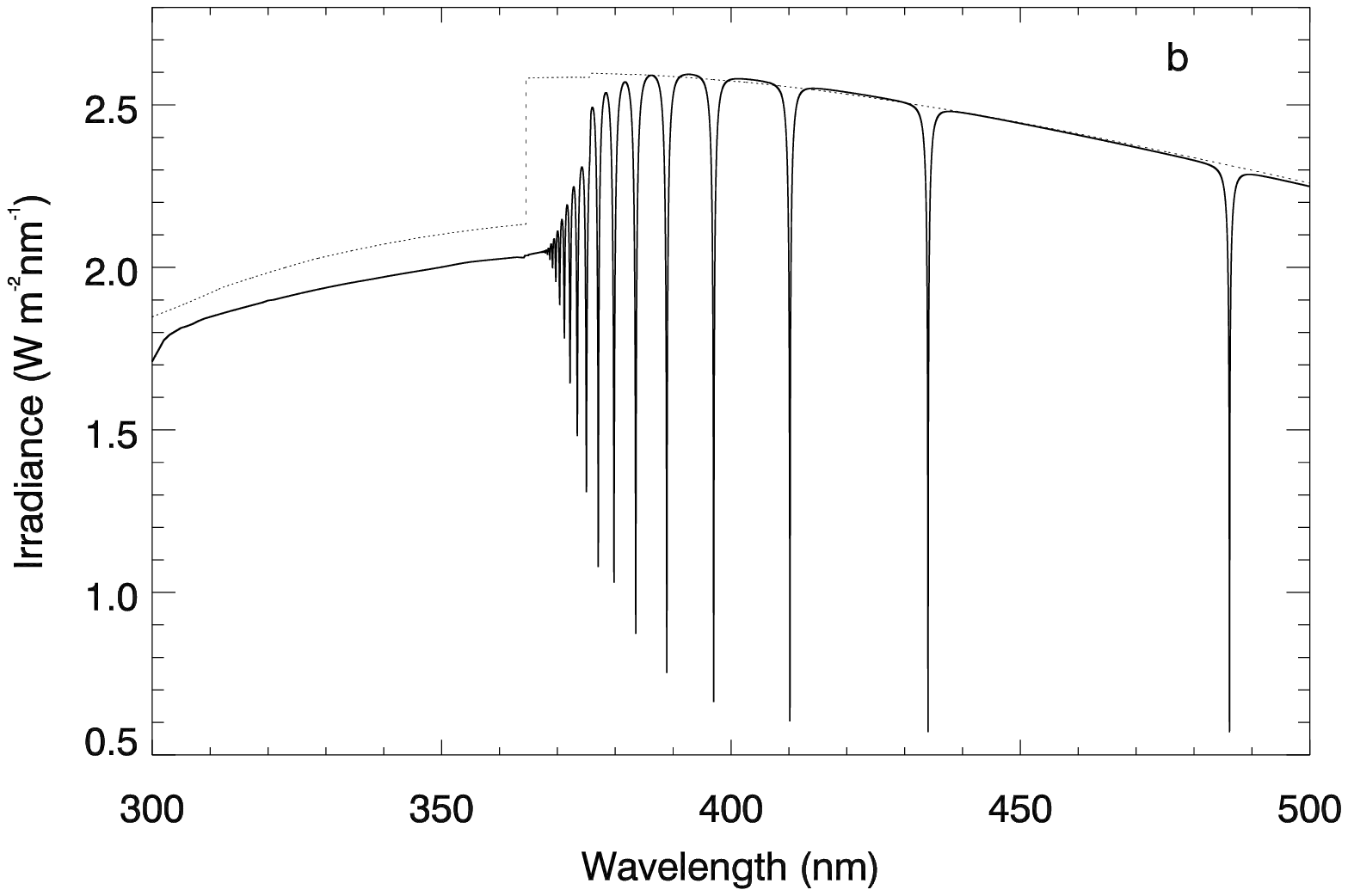}
\caption{As Fig.\,\ref{fig:LTEIP0}a for the wavelength range 200 to 300\,nm (panel a), and 300 to 500\,nm (panel b). The differences are due to different photoionization cross sections implemented in COSI. The lines in panel (b) are the hydrogen lines longward of the Balmer jump. 
\label{fig:atl_cosi}}  
\end{figure*}

\begin{figure*}
\centering
\includegraphics[width=.49\linewidth]{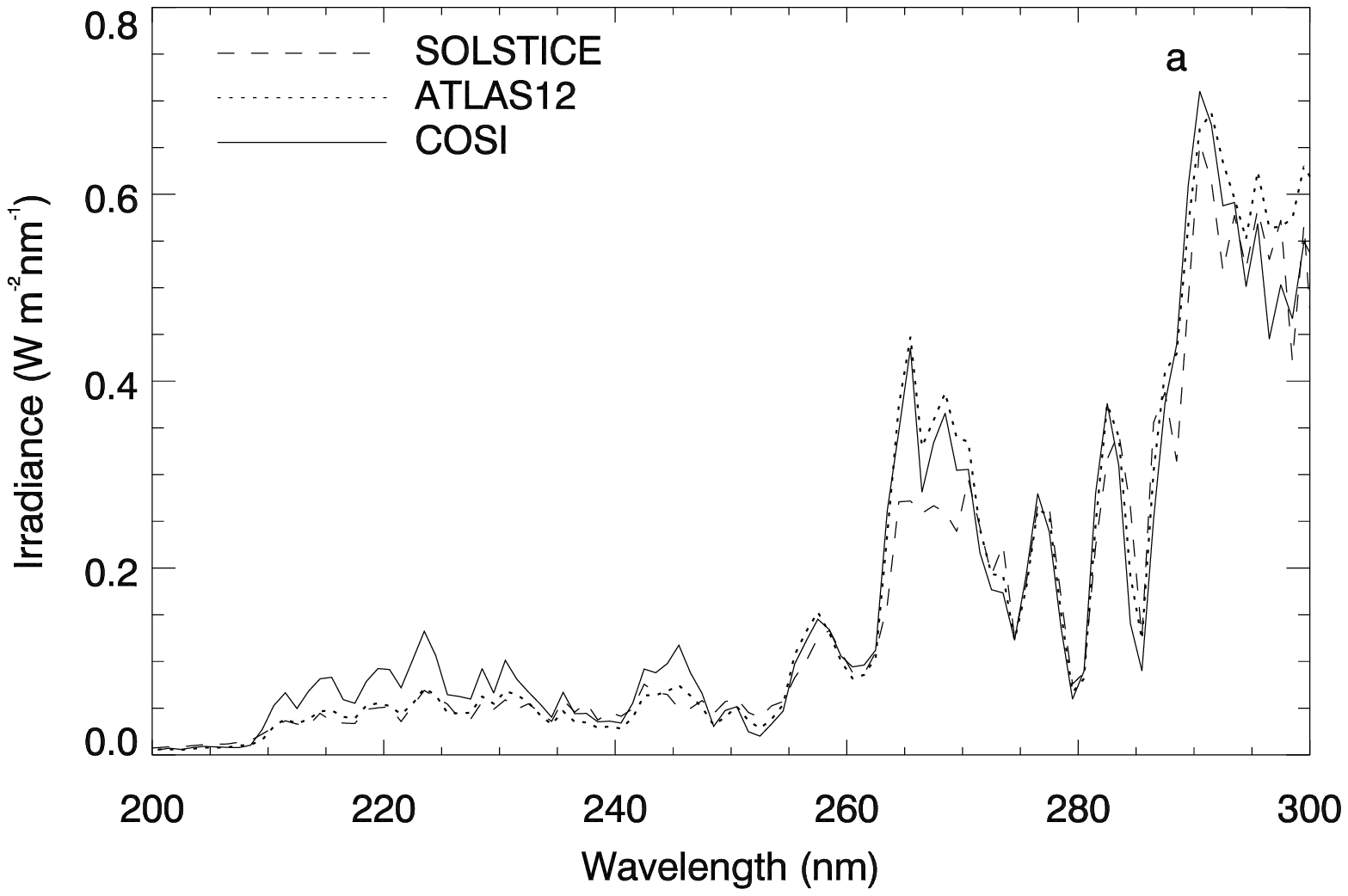}
\includegraphics[width=.49\linewidth]{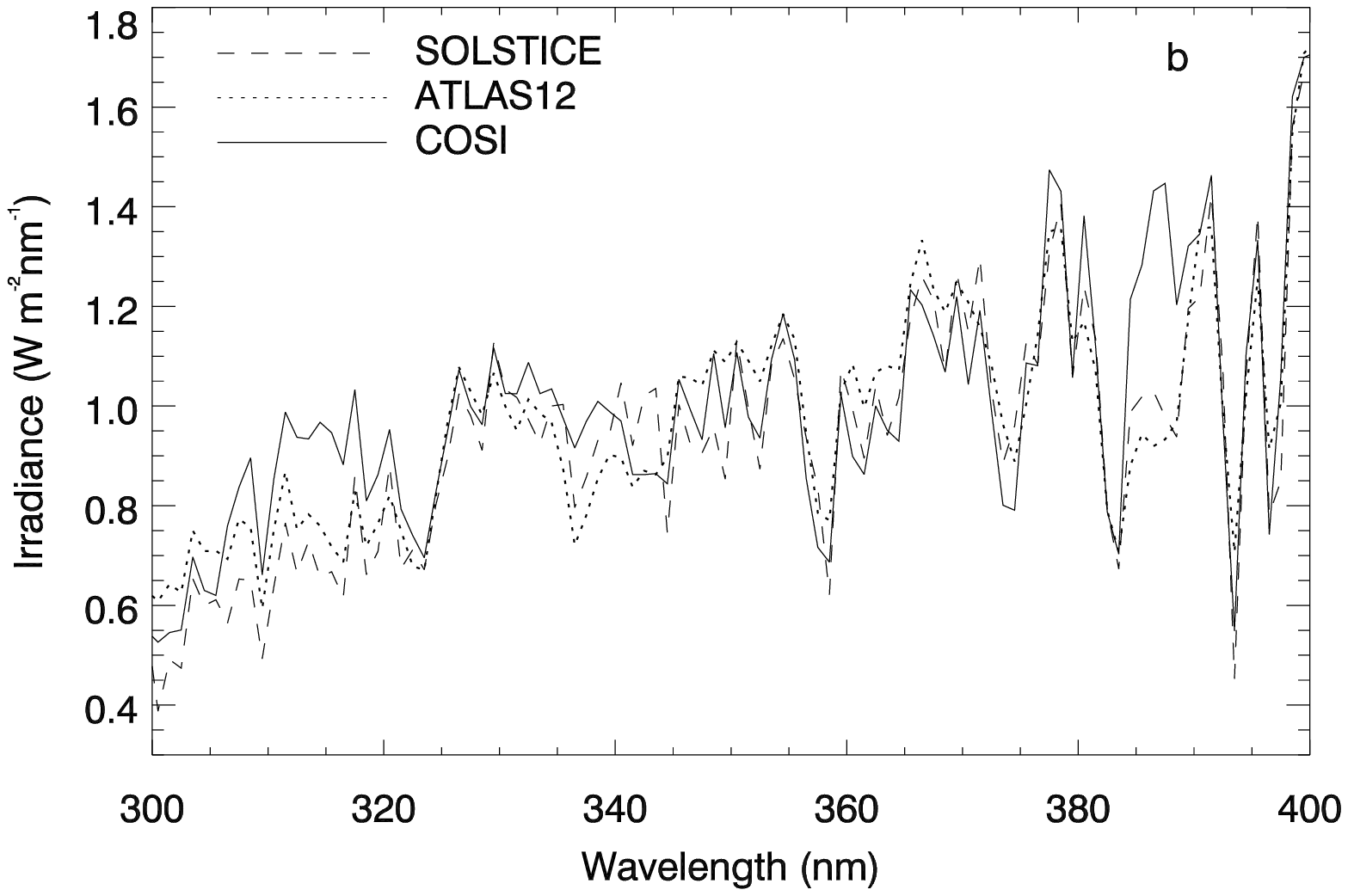}
\caption{As Fig.\,\ref{fig:LTEIP0}b plus a comparison with the SOLSTICE measurements (dashed line). The synthetic spectra are convolved with 1-nm boxcar filters to reproduce the resolution of the observation. The higher flux between 210 and 235\,nm can be explained by missing continuum and line opacity. The discrepancy between 385 and 390\,nm is due to the CN-band not calculated with COSI. 
\label{fig:NLTEsolstATLAS}} 
\end{figure*}

To further illustrate the COSI to ATLAS12 comparison, Fig.\,\ref{fig:atl_cosi} shows the same continuum calculation as in Fig.\,\ref{fig:LTEIP0}a, but over a narrower wavelength range in the UV. The discrepancy between 210 and 250\,nm (panel a) can be explained by different photoionization cross sections which lead to a higher continuum flux in the COSI calculation than in the ATLAS12 calculation. We tested the effect of increasing the Mg\,{\sc{i}}\,2 photoionization cross section by a factor of 2 in COSI. This leads to a much better agreement between both calculations but is inconsistent with Opacity Project data. The lower continuum calculated with COSI between 290 and 300\,nm is a consequence of a resonance in the photoionization cross section of Mg\,{\sc{i}} 3 level (cf. Fig.\,\ref{fig:cs}). However, this resonance is not confirmed by the observations taken with the Solar Stellar Irradiance Comparison Experiment (SOLSTICE) \citep{Rottman1993,Woods1993}, as the measurements agree better with the ATLAS12 calculations. This indicates that there is still a need for improved atomic data. Nevertheless, from 260 and 280\,nm (panel a) the continua give comparable results. In Fig.\,\ref{fig:NLTEsolstATLAS} the LTE line spectra calculated with COSI are validated against the SOLSTICE measurements. The fairly good agreement indicates that the spectrum synthesis with COSI in LTE is calculated reasonably well. However, from $\sim$310 to 320\,nm COSI leads to a considerable higher irradiance than ATLAS12 and the SOLSTICE observation. The missing molecular lines of the CN-band explain the discrepancy between $\sim$385 and 390\,nm. 
\section{NLTE calculations}\label{sec:nlte}
\subsection{NLTE-ODFs}
\subsubsection{Theory}
For a complete calculation of the radiative transfer all rele\-vant processes, e.g. the bound-bound, bound-free, and free-free processes have to be taken into account. Due to the huge number of bound-bound transitions the opa\-city of all spectral lines is considerable and must be taken into account in the NLTE radiative transfer calculation. In COSI only the line transitions between the explicit levels given in Table\,\ref{tab:levels} are calculated in NLTE. Thus, the line opacity of a large number of lines is not directly included in the solution of the NLTE radiative transport. Indirectly we account for them by applying opacity distribution functions (ODFs) of the line opacities. The concept of ODFs has been first suggested by \cite{StromKurucz1966}, for a review see \cite{Carbon1984}. To our knowledge, this concept has not yet been used with an iterative scheme in a NLTE calculation. 
\subsubsection{Implementation of NLTE-ODFs}\label{sec:lbkg}  
\begin{figure}
\centering
\includegraphics[width=\linewidth]{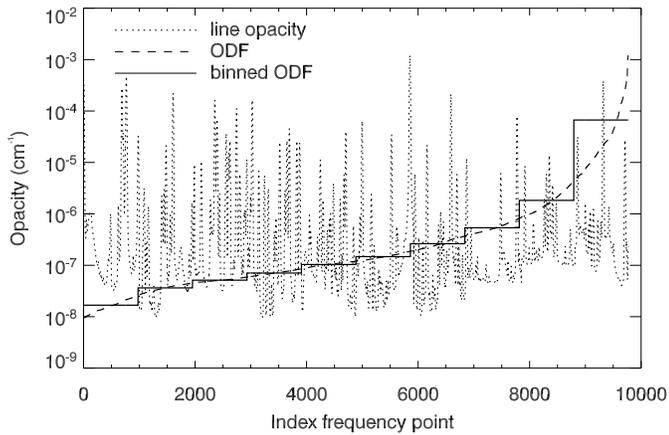}
\caption{Line opacities for the wavelength range from 200 to 210\,nm for the depth point where $\tau_{\mathrm{500}}$=1. The huge number of frequency points necessary to account for the detailed line opacities is reduced to 10 values by means of the ODFs. This allows the inclusion of the line opacities into the NLTE radiative transfer with a reasonable computational effort. \label{fig:line_op}}
\end{figure} 
{\b For the calculation of the ODFs the high resolution line opacities need to be calculated over the wavelength range under consideration, which in our case is from 90 to 400\,nm.} The computing time would be prohibitive to account for each single line in the NLTE solution and therefore we describe all the line opacities by distribution functions over a certain wavelength range $\Delta \lambda$ (in our case 10\,nm) as a function of line strength. This means that over $\Delta \lambda$ and for each depth point the line opacities are sorted according to their strengths. Then, for a certain number of opacity bins (in our case 10) we calculate the mean opacity $\bar{\kappa}_j$ for each bin $j$ as given in Eq.\,\ref{eq:odf}. 
\begin{equation}
\mathrm{mean}(\mathrm{sort}(\kappa_i))_j \rightarrow \bar{\kappa}_j \label{eq:odf}
\end{equation}
with $i$ being the frequency index of the high resolution opacities. The bins $j$ are then transfered to evenly spaced wavelength points within the wavelength interval under consideration yielding NLTE-ODFs with a 1\,nm resolution. Fig.\,\ref{fig:line_op} shows the high resolution line opacities (dotted line) from 200 to 210\,nm at the depth where $\tau_{\mathrm{500}}$= 1. Also shown are the calculated ODF (dashed line) and the binned mean opacity (solid line) again for the same depth point. Thus, we include values of averaged line opacity, representing the high resolution line opacity, in the NLTE calculation. 

To avoid the excess of the chromospheric emission for lines that are calculated in LTE, the line emissivity for layers outward the temperature minimum is set to the value calculated from the Planck function at the temperature minimum. We are aware that this approach clearly underestimates the emissivity for many upper chromosphere lines and it will be improved in the future through the inclusion of more levels in the NLTE calculation and/or by an approximative treatment of the non-LTE effects, e.g. using the modified excitation temperatures as introduced by \cite{Schmutz1991}.

Although the NLTE-ODF distribution function is calculated from LTE level populations there is an indirect effect of the NLTE population through changes of the ionization equilibrium, i.e. all of the level populations that are treated in LTE with respect to the NLTE ground state population (cf. Table\,\ref{tab:levels}) are indirectly affected by the NLTE effects. Thus the inclusion of the ODFs changes the line strengths of the emergent spectrum, which in turn leads to a different set of ODFs. Therefore, self-consistent ODFs can only be calculated iteratively. We have investigated to what extent a set of successive iterations of ODFs influences the population values of the explicit NLTE levels. We also studied how many iterations are necessary to find a solution for which the ODFs are practically unchanged in a next iteration. The changes of the population numbers calculated without and with the first set of ODFs are up to a factor 30. The next iteration leads to changes of 1.5, the second and third iterations show maximal changes of a factor of 1.002. As the population numbers are practically unchanged between the second and third set of ODFs, we conclude that two iterations are sufficient to find a self-consistent distribution of ODFs \citep{Haberreiter2006PhDT}.  

\begin{figure*}
\centering
\includegraphics[width=0.49\linewidth]{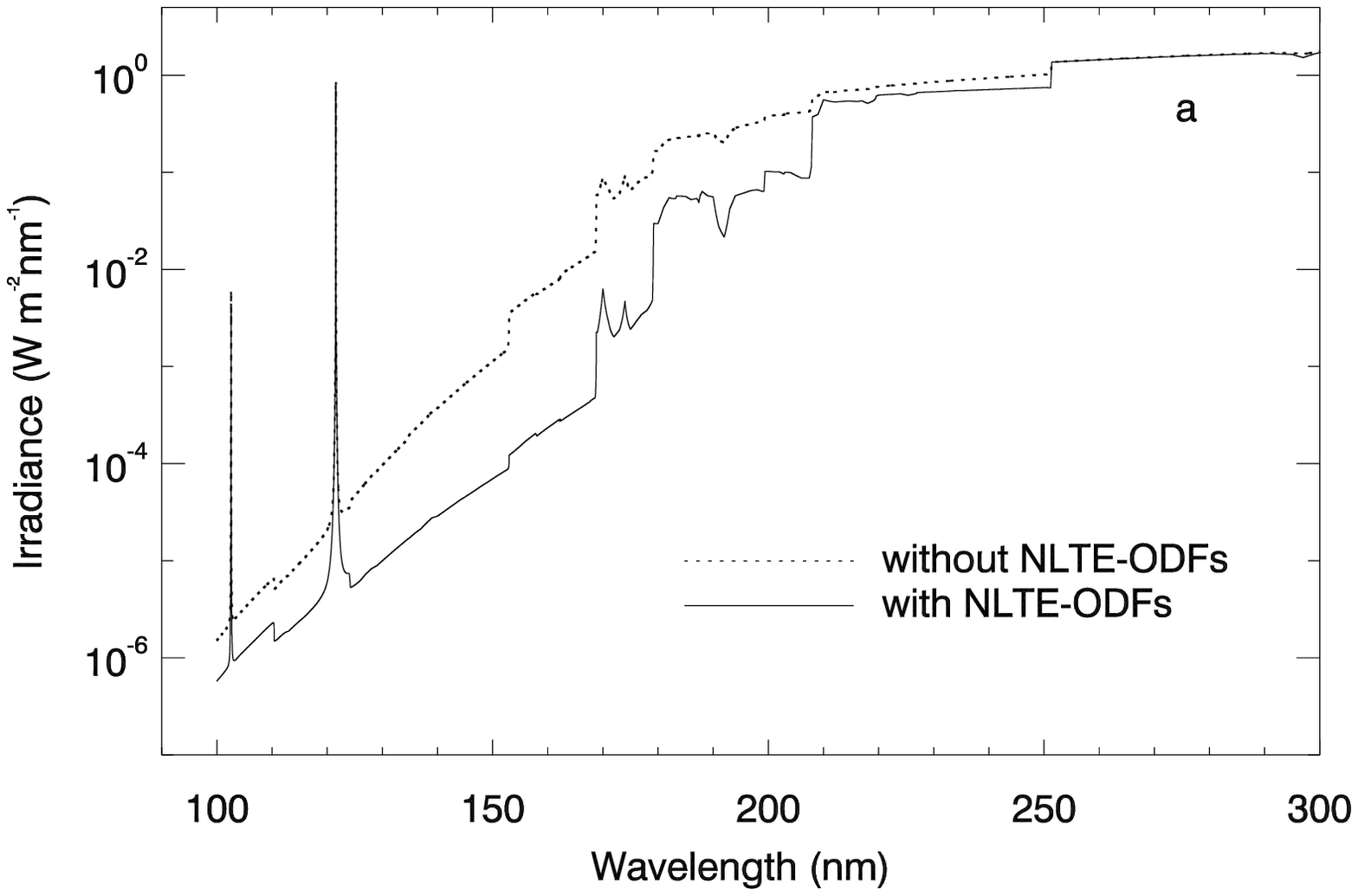}
\includegraphics[width=0.49\linewidth]{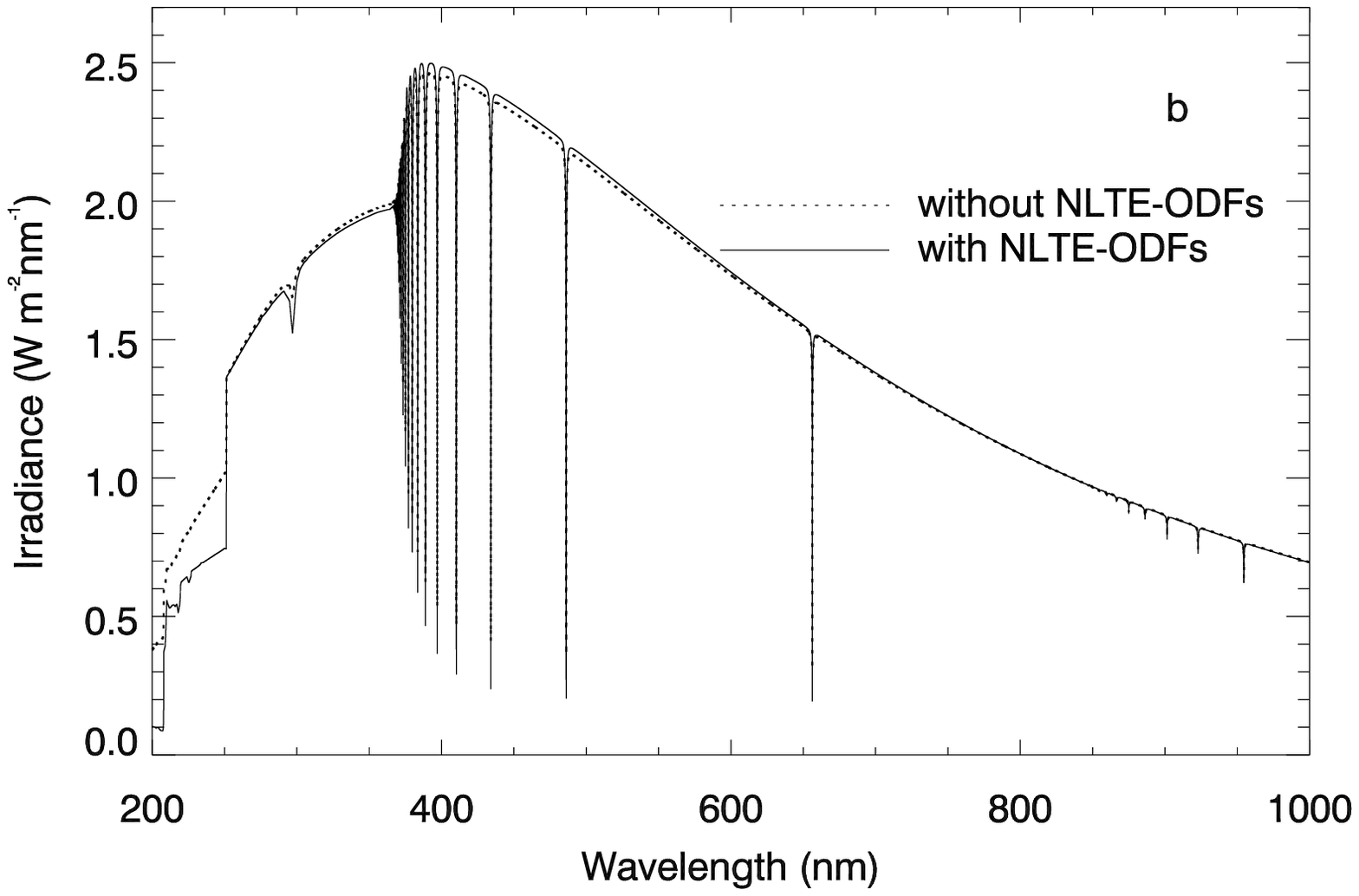}
\caption{Comparison of the continuum calculations carried out with COSI in NLTE and using the F1999 model atmosphere structure to analyze the effect of the NLTE-ODFs on the calculated continuum. The implemented NLTE-ODFs for 100 to 400\,nm lead to a decrease of the continuum flux, which in turn leads to less ionization. The decrease in electron density leads to a reduction of negative hydrogen opacity and therefore to a slight increase in the visible continuum (panel b). \label{fig:cosi_cont_odf}}  
\end{figure*}
\begin{figure*}
\centering
\includegraphics[width=.49\linewidth]{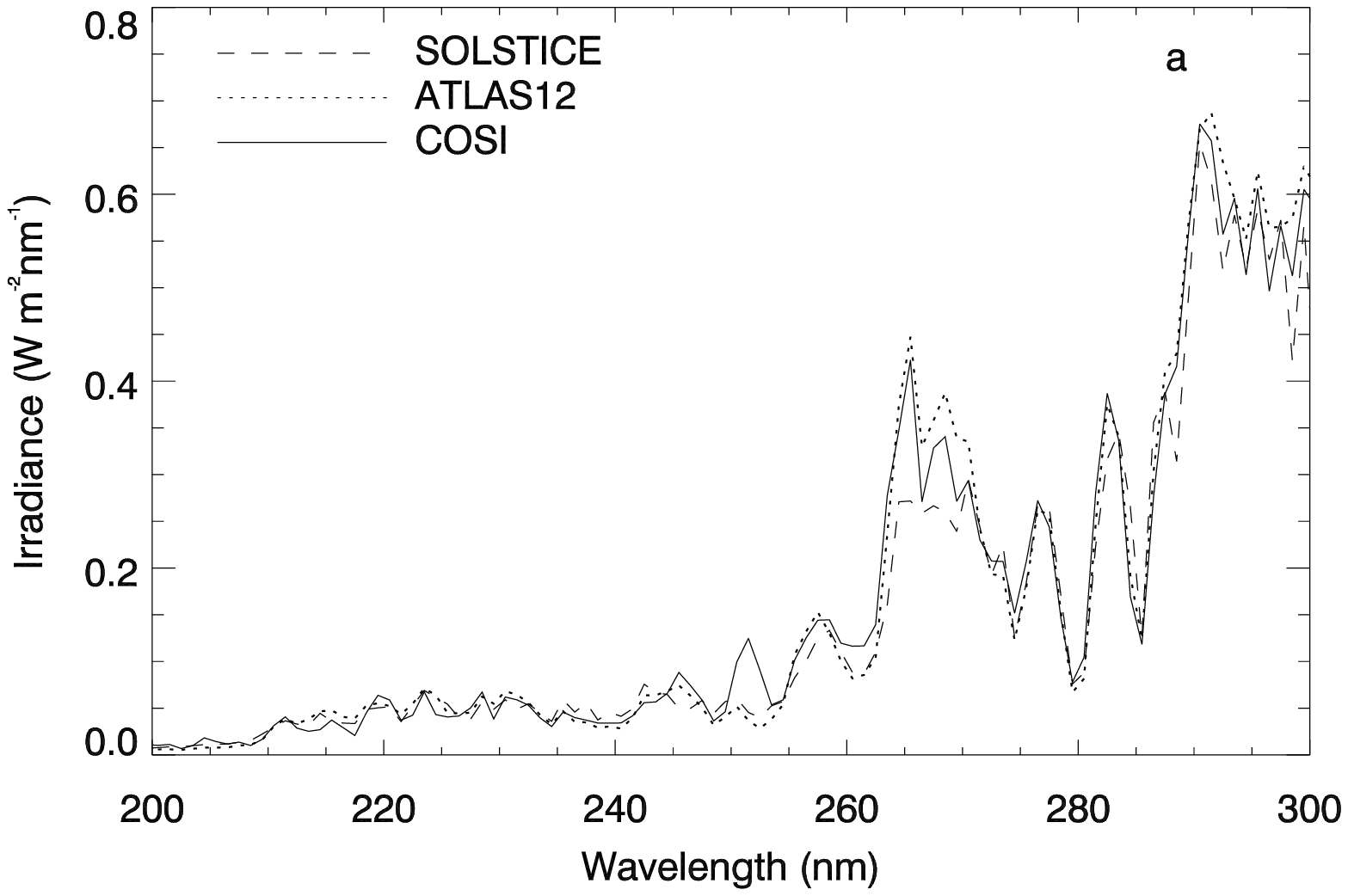}
\includegraphics[width=.49\linewidth]{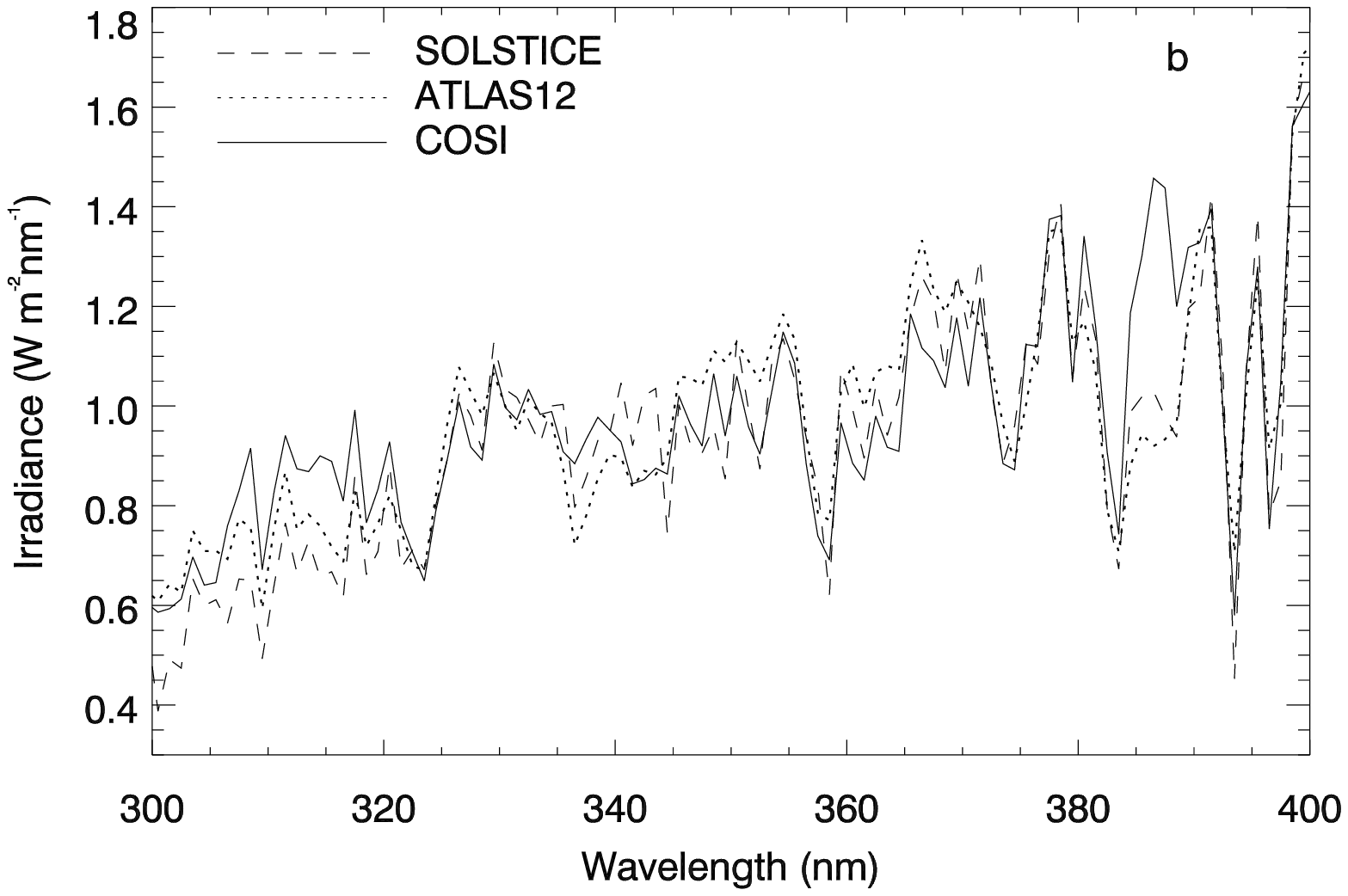}
\caption{Comparison of the synthetic spectrum calculated with COSI in NLTE using F1999 as atmosphere structure with the SOLSTICE measurements (dashed line) and the ATLAS12 calculation. The synthetic spectra are convolved with a 1-nm boxcar to reproduce the resolution of the observation. \label{fig:solstNLTE}} 
\end{figure*}
\begin{figure}
\centering
\includegraphics[width=1.\linewidth]{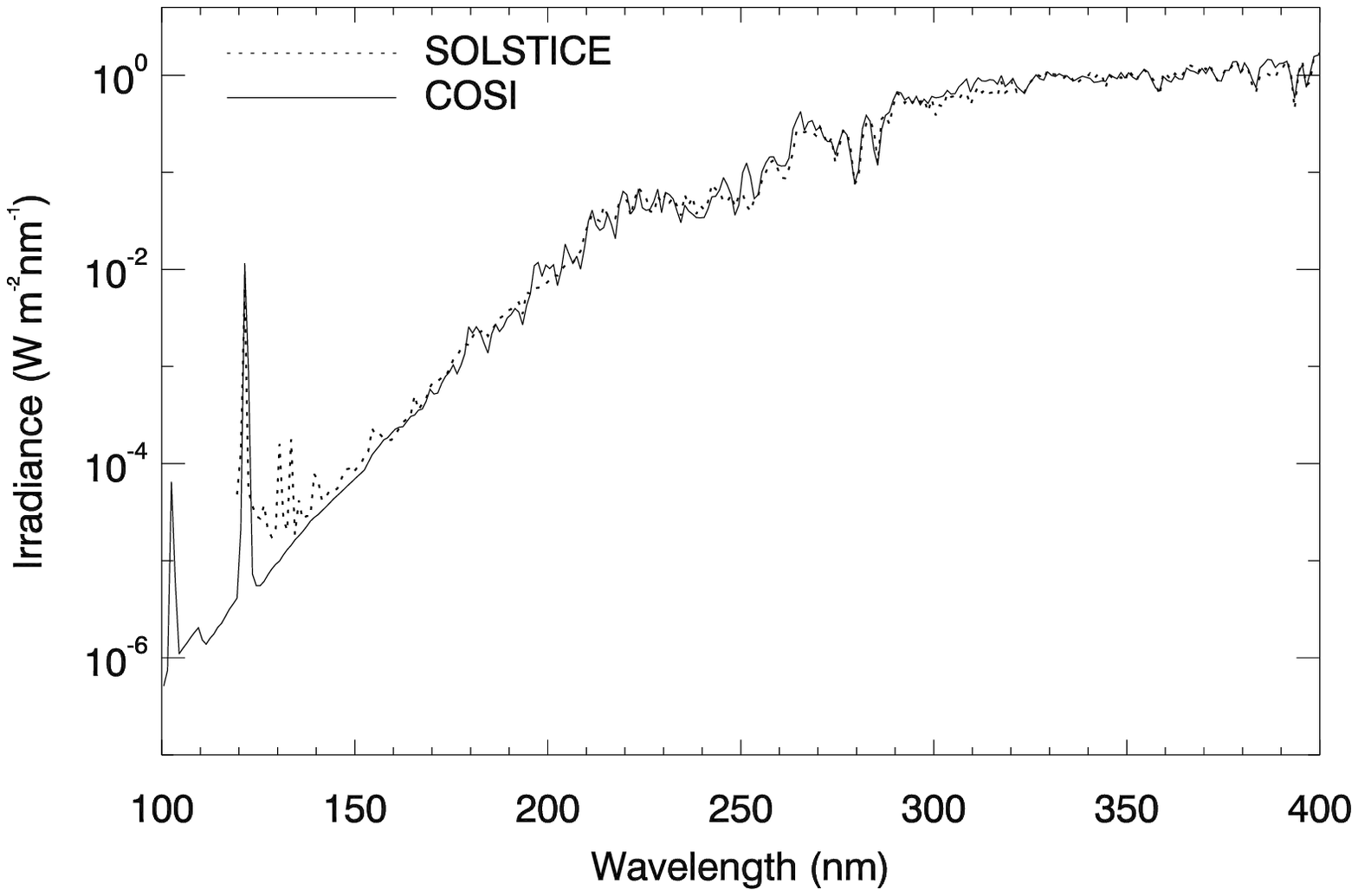}
\caption{Comparison of the synthetic spectrum calculated with COSI in NLTE with the F1999 atmosphere structure (solid line) together with the SOLSTICE observation at solar minimum on April 3, 1997 (dashed line). An increased line broa\-dening has been applied to account for missing opacities.  \label{fig:solstNLTEbroad}} 
\end{figure}

\subsubsection{Results with NLTE-ODFs}
In this section we present NLTE calculations carried out with COSI and using the F1999 model by \cite{Fontenla1999} along with the F2007 abundance values, and the NLTE-ODFs as described in the section above. In Fig.\,\ref{fig:cosi_cont_odf} the effect of the NLTE-ODFs on the emergent continuum spectrum is shown. The dotted line represents the continuum spectrum calculated without NLTE-ODFs, the solid line shows the spectrum with NLTE-ODFs. The implementation of the NLTE-ODFs leads to a considerable decrease of the flux from $\sim$100 to 200\,nm. This is due to the fact that the NLTE-ODFs increase the opacity. Consequently, the height where $\tau$ equals 1 is higher up in the solar atmosphere where the temperature is lower than in the case for the pure continuum opacity. Thus, the UV flux is decreased and the elements, in particular the metals, are less ionized. The increase of neutral atoms in turn leads to an increase of the continuum opacity and thus to a decrease of the continuum flux. It is important to note that the negative hydrogen opacity strongly depends on the electron density. As the ionization decreases, less negative hydrogen is formed due to a decrease in the electron density in some layers, leading to a slight decrease of the continuum flux in the optical wavelength range (Fig\,\ref{fig:cosi_cont_odf}b). Note that COSI calculates the negative hydrogen in full NLTE, i.e.\ in a simultaneous solution together with all other NLTE levels as given in Table\,\ref{tab:levels}.

Fig.\,\ref{fig:solstNLTE} shows the corresponding line spectrum calculated in NLTE with COSI together with SOLSTICE measurements and ATLAS12 calculation. In order to compensate for the missing opacities, which is also partly present in the COSI LTE calculation, the Doppler broadening was increased from 1.5\,km/s to 15\,km/s in the wavelength range from 170 to 210\,nm, from 1.5\,km/s to 8\,km/s from 210 to 250\,nm and 1.5\,km/s to 3 km/s from 250 to 400\,nm. For the wavelength range from 300 to 400\,nm the COSI calculation is in reasonable agreement with the observation. To reproduce the observation it is clear that the implementation of the NLTE-ODFs is important, because without them the emergent flux would be considerably higher, as already shown in Fig.\,\ref{fig:cosi_cont_odf}. Nevertheless, the remaining issues about the missing opacities need further investigation. For future calculations the inclusion of molecular lines and additional continuum opacities such as CH, OH and CN photodissociation are necessary.

Finally, Fig.\,\ref{fig:solstNLTEbroad} shows the comparison of the line spectrum calculated in NLTE with COSI together with SOLSTICE measurements for the wavelength range from 100 to 400\,nm. The overall agreement between the calculation and the observation is good. However, the continuum around the Lyman-$\alpha$ line is considerably lower than the observation. One of the reasons for this is clearly the missing contribution of the quiet Sun network and also enhanced network to the calculated spectrum. Also, the fact that our approach underestimates the emissivity of the chromospheric emission lines, can further explain the discrepancy below 150\,nm. Nevertheless, the general agreement above 150\,nm, in particular the continuum slope, clearly shows that the inclusion of the line opacities is essential for the calculation of the solar spectrum in the UV.
\section{Discussion\label{sec:disc}}
In the present version of COSI we have calculated the NLTE-ODFs with line opacities in LTE with respect to the ground state of the ion. For all heights above the temperature minimum the emissivity is set to $\epsilon_{L,\lambda}=B_{\lambda, T_{\mathrm{min}}}\kappa_{L,\lambda}$ with $B$ being the Planck function and $L$ the depth point index of the atmosphere structure. In F1999 the layers around $T_{\mathrm{min}}$ are in near-LTE conditions which makes our approach suitable for this and similar stellar atmosphere models. 
However, if we would calculate an atmosphere with strong departures from LTE, as e.g. SRPM\,305, our assumptions would lead to a substantial overestimation of the opacity and at the same time an underestimation of the source function above the temperature minimum, which in turn would lead to a much lower UV continuum intensity than in the observations shown in the calculations by \cite{Fontenla2007}. In order to address such structures the NLTE effects have to be accounted for in more detail.

\cite{Hab2005AdSpR} presented a comparison of the calculated Lyman-$\alpha$ line strength as a function of time over two solar cycles with its observed strength as compiled by \cite{Woods2000}. Their calculations are based on the same F1999 COSI model as discussed above. They found that the synthetic Lyman-$\alpha$ line is about a factor of 2 too weak. The discrepancy has been found in part to be due to ambipolar diffusion of protons and neutral hydrogen as implemented in F1999 and the later atmosphere structure SRPM305 by \cite{Fontenla2007} but not yet in the COSI code. Another important contribution which leads to an increased Lyman\,$\alpha$ line strength is the expansion of the area of active regions when a line is formed higher up in the solar atmosphere \citep{Schoell2008}. 

\section{Conclusions}\label{sec:concl}
We presented the NLTE radiative transfer code COSI for the calculation of the spectral solar energy distribution and validated its performance in LTE against ATLAS12 calculations and observed solar spectra. The results show that COSI calculates correctly the spectra in LTE. Furthermore, we have introduced the new concept of NLTE-ODFs. This new technique allows all line opacities to be included into the NLTE radiative transfer. The fast convergence of the NLTE-ODFs does not significantly increase the computing time. 
We conclude that by inclu\-ding NLTE-ODFs into the radiative transfer code COSI the calculations of the solar spectra are improved substantially. It is important to note that this new concept is not limited to specific atoms, and can in principle also be used in an iterative scheme to derive stellar atmosphere structures.
\begin{acknowledgements}
This work was supported through ETH Polyproject {\em{Variability of the Sun and Global Climate}} and SNF Project No. 200020-109420. We would like to give special thanks to Robert Kurucz for providing the synthetic spectra calculated with ATLAS12, as well as the line opa\-ci\-ties. We very warmly acknowledge that he kindly answered the many questions we had concerning the solar spectra. We are also grateful to Juan Fontenla for providing the atmosphere structures and many fruitful discussions. 
\end{acknowledgements}

\clearpage \onecolumn 
\begin{appendix}
\section{Atomic model}
\begin{table*}[!hh]
\caption{Levels of the atomic model that are explicitly treated in NLTE by COSI. Given are the ion, the level and the atomic numbers AN; electron configuration and spectroscopic term designation, the ionization wavelength $\lambda_{\mathrm{ion}}$ calculated from the measured energies from the NIST database, where * denotes ionization to an excited state of the next higher ion; the ionization potential $E_{\mathrm{ion}}$ of the ground state; the energy $E$ of the level, which is the weighted mean in the case of split levels; the statistical weight $g$ of the model configuration. Finally the last column denotes the source from which the photoionization cross sections were taken. Here J88 refers to \cite{John1988}, MIH to \cite{Mihalas1967}, SEAT to \cite{Seaton1994} and KOE to \cite{Koester1985}. Furthermore, {\em Hyd} stands for the hydrogenic treatment and {\em OP} means that the photoionization cross sections are from the OPACITY PROJECT and in case of iron from the IRON PROJECT. \label{tab:levels}}  
\begin{center}
{\normalsize{
    \begin{tabular*}{.85\textwidth}{lrrllrrrrc}
      \hline
      \hline
      {{Ion}}  & 
      {{Level}}  &
      {{AN}}  & 
      {{Configuration}} &
      {{Term}} & 
      {{$\lambda_{\mathrm{ion}}$ (\AA)}}&  
      {{$E_{\mathrm{ion}}$} ($\mathrm{cm^{-1}}$)} &
      {{$E$} ($\mathrm{cm^{-1}}$)}&
      {{${g}$}}&
      {{source}}\\  
      \hline
\noalign{\smallskip}
     H$^{-}$ & 1&1& $\mathrm{1s}$      & $\mathrm{^{1}S}$ & 16414.54        & 6090.50 & 0.0 	& 2 & J88 \\
     H\,{\sc{i}} & 1&1& $\mathrm{1s}$  & $\mathrm{^{2}S}$         &   911.77   & 109677.38 & 0.0 	& 2 & MIH \\
     H\,{\sc{i}} & 2&1& $\mathrm{2p}$  & $\mathrm{^{2}P^{\circ}}$ &  3645.49   &  & 82258.04 	& 8 & MIH \\
     H\,{\sc{i}} & 3&1& $\mathrm{2s}$  & $\mathrm{^{2}S}$         &  8200.65   &  & 97491.00	& 18 & MIH \\
     H\,{\sc{i}} & 4&1& $\mathrm{3p}$  & $\mathrm{^{2}P^{\circ}}$ & 14576.68   &  & 102822.54	& 32 & MIH \\
     H\,{\sc{i}} & 5&1& $\mathrm{3s}$  & $\mathrm{^{2}S}$         & 22773.63   &  & 105290.28       & 50 & MIH \\
     H\,{\sc{i}} & 6&1& $\mathrm{3d}$  & $\mathrm{^{2}D}$         & 32791.60   &  & 106630.79	& 72 & MIH \\
     H\,{\sc{i}} & 7&1& $\mathrm{4p}$  & $\mathrm{^{2}P^{\circ}}$ & 44630.70   &  & 107439.07	& 98 & MIH \\
     H\,{\sc{i}} & 8&1& $\mathrm{4s}$  & $\mathrm{^{2}S}$         & 58291.00   &  & 107963.67	& 128 & MIH \\
     H\,{\sc{i}} & 9&1& $\mathrm{4d}$  & $\mathrm{^{2}D}$         & 73772.54   &  & 108323.34 	& 162 & MIH \\
     H\,{\sc{i}} & 10&1& $\mathrm{4f}$ & $\mathrm{^{2}P^{\circ}}$ & 91075.35   &  & 108580.61	& 200 & MIH \\
     H\,{\sc{ii}} & 1&1& -       & -  &         &  & 0.0 	& 1 & - \\
    He\,{\sc{i}} & 1&2& $\mathrm{1s^2}$      & $\mathrm{^{1}S}$ &   504.25    				& 8200.65  & 0.0 	& 1 &  KOE\\
    He\,{\sc{i}} & 2&2& $\mathrm{1s2s}$      & $\mathrm{^{3}S}$ &  2601.00    				&  & 159856.07	& 3 &  KOE\\
    He\,{\sc{i}} & 3&2& $\mathrm{1s2s}$      & $\mathrm{^{1}S}$ &  3122.27    				&  & 166277.55	& 1 &  KOE\\
    He\,{\sc{i}} & 4&2& $\mathrm{1s2p}$      & $\mathrm{^{3}P^{\circ}}$ &  3422.37    &  & 169086.94 	& 7 &  KOE\\
    He\,{\sc{i}} & 5&2& $\mathrm{1s2p}$      & $\mathrm{^{1}P^{\circ}}$ &  3680.26    &  & 171135.00	& 3 &  KOE\\
    He\,{\sc{i}} & 6&2& $\mathrm{1s3s}$      & $\mathrm{^{3}S}$ &  6634.61    &  & 183236.89	& 3 &  KOE\\
    He\,{\sc{i}} & 7&2& $\mathrm{1s3s}$      & $\mathrm{^{1}S}$ &  7437.69    &  & 184864.55	& 1 &  KOE\\
    He\,{\sc{i}} & 8&2& $\mathrm{1s3p}$      & $\mathrm{^{3}P^{\circ}}$ &  7846.22    &  & 185564.68	& 7 &  KOE\\
    He\,{\sc{i}} & 9&2& $\mathrm{1s3d}$      & $\mathrm{^{3}D}$ &  8191.29    &  & 186101.65	& 13 &  KOE\\
    He\,{\sc{i}} & 10&2& $\mathrm{1s3d}$     & $\mathrm{^{1}D}$ &  8193.59    &  & 186105.07  & 5 &  KOE\\

    He\,{\sc{ii}} & 1&2& $\mathrm{1s}$       & $\mathrm{^{2}S}$ &        &   438908.85& 0.0 		& 2 &  -\\

     Li\,{\sc{i}} & 1&3& [He]$\mathrm{2s}$   & $\mathrm{^{2}S}$ &   2298.90      & 43487.15& 0.0 	& 2 & Hyd \\
     Li\,{\sc{ii}}& 1&3& [He]        	       & $\mathrm{^{1}S}$ &       &       & 0.0 		& 1 & - \\
     Be\,{\sc{i}} & 1&4& [He]$\mathrm{2s^2}$ & $\mathrm{^{1}S}$ &   1329.92     & 75192.64& 0.0 	& 1 & Hyd\\
     Be\,{\sc{ii}}& 1&4& [He]$\mathrm{2s}$   & $\mathrm{^{2}S}$ &               &        & 0.0 	& 2 &  - \\
     B\,{\sc{i}}  & 1&5& [He]$\mathrm{2s^2 2p}$ & $\mathrm{^{2}P^{\circ}}$ & 1494.14 & 66928.10& 0.0 	& 2 & Hyd\\
     B\,{\sc{ii}} & 1&5& [He]$\mathrm{2s^2}$    & $\mathrm{^{1}S}$ &  & & 0.0 			& 1 &  - \\
     C\,{\sc{i}} &1&6& [He]$\mathrm{2s^2 2p^2}$ & $\mathrm{^{3}P}$ & 1101.79& 90820.42&  29.57	& 9 & OP\\
     C\,{\sc{i}} &2&6& [He]$\mathrm{2s^2 2p^2}$ & $\mathrm{^{1}D}$ & 1240.72& 			& 10192.63 & 5& OP\\
     C\,{\sc{i}} &3&6& [He]$\mathrm{2s^2 2p^2}$ & $\mathrm{^{1}S}$ & 1446.28&			& 21648.01 & 1& OP\\       
     C\,{\sc{ii}} &1& 6&[He]$\mathrm{2s^2 2p}$& $\mathrm{^{2}P^{\circ}}$&	&		& 0.00 &6& - \\ 
     N\,{\sc{i}} &1&7& [He]$\mathrm{2s^2 2p^3}$  & $\mathrm{^{4}S^{\circ}}$ & 852.62 & 117314.58& 0.0 & 4& Hyd \\
     N\,{\sc{ii}}&1&7& [He]$\mathrm{2s^2 2p^2}$  & $\mathrm{^{3}P}$         &  & & 0.0 		        & 9 & - \\
     O\,{\sc{i}} &1&8& [He]$\mathrm{2s^2 2p^4}$  & $\mathrm{^{3}P}$         &  911.73& 109837.02& 77.97 	& 9 &  Hyd\\
     O\,{\sc{ii}}&1&8& [He]$\mathrm{2s^2 2p^3}$ & $\mathrm{^{4}S^{\circ}}$  &  & & 0.0 		& 4 & - \\
     F\,{\sc{i}} &1&9& [He]$\mathrm{2s^2 2p^5}$  & $\mathrm{^{3}P^{\circ}}$ &  712.99& 140524.5 & 134.7  & 6 & Hyd\\
     F\,{\sc{ii}}&1&9& [He]$\mathrm{2s^2 2p^4}$ & $\mathrm{^{3}P}$          &  & & 0.0 			& 9 & - \\
     Ne\,{\sc{i}} &1&10& [He]$\mathrm{2s^2 2p^6}$  & $\mathrm{^{1}S}$       &  574.94&  173929.75&  	& 1 &  Hyd\\
     Ne\,{\sc{ii}}&1&10& [He]$\mathrm{2s^2 2p^5}$ & $\mathrm{^{2}P^{\circ}}$ &  & & 0.0 	& 6 & -\\
		      \noalign{\smallskip}
		      \hline
		  \end{tabular*}
		  }}
\end{center}
\end{table*}
\begin{table*}[!hh]  
\begin{center}
\addtocounter{table}{-1}
{\normalsize{
    \begin{tabular*}{.85\textwidth}{lrrllrrrrc}
      \hline
      \hline
      {{Ion}}  & 
      {{Level}}  &
      {{AN}}  & 
      {{Configuration}} &
      {{Term}} & 
      {{$\lambda_{\mathrm{ion}}$ (\AA)}}&  
      {{$E_{\mathrm{ion}}$} ($\mathrm{cm^{-1}}$)} &
      {{$\bar{E}$} ($\mathrm{cm^{-1}}$)}&
      {{$\bar{g}$}}&
      {{cross}}\\  
      \hline
\noalign{\smallskip}
			Na\,{\sc{i}} &1&11&[Ne]$\mathrm{3s}$  & $\mathrm{^{2}S}$&2411.91&41449.62& 0.00 &2 & OP\\
			Na\,{\sc{i}} &2&11&[Ne]$\mathrm{3p}$  & $\mathrm{^{2}P^{\circ}}$&4083.51& & 16967.64&6 & OP\\
			Na\,{\sc{i}} &3&11&[Ne]$\mathrm{4s}$   & $\mathrm{^{2}S}$&6363.72 & & 25739.99&2 & OP\\ 
			Na\,{\sc{i}} &4&11&[Ne]$\mathrm{3d}$   & $\mathrm{^{2}D}$& 8143.14& & 29172.86&10 & OP\\ 
			Na\,{\sc{i}} &5&11&[Ne]$\mathrm{4p}$   & $\mathrm{^{2}P^{\circ}}$&8942.85 & & 30270.72&6 & OP\\ 
			Na\,{\sc{i}} &6&11&[Ne]$\mathrm{5s}$   & $\mathrm{^{2}S}$&12119.24 & & 33200.68&2 & OP\\ 
			Na\,{\sc{i}} &7&11&[Ne]$\mathrm{4d}$   & $\mathrm{^{2}D}$&14486.66 & & 34548.74&10 & OP\\ 
			Na\,{\sc{i}} &8&11&[Ne]$\mathrm{4f}$   & $\mathrm{^{2}F^{\circ}}$&14567.25 & & 34586.92&14 & OP\\ 
			Na\,{\sc{i}} &9&11&[Ne]$\mathrm{5p}$   & $\mathrm{^{2}P^{\circ}}$&15601.91 & & 35042.03&6 & OP\\ 
			Na\,{\sc{i}} &10&11&[Ne]$\mathrm{6s}$  & $\mathrm{^{2}S}$&19690.85  & & 36372.62&2 & OP\\ 
      Na\,{\sc{ii}} &1 & 11&[Ne]          &  $\mathrm{^{1}S}$&  & & 0.00 &1& -\\    
     Mg\,{\sc{i}} &1& 12&[Ne]$\mathrm{3s^2}$& $\mathrm{^{1}S}$	&  1621.51&  61671.02&0.00 &1 & OP\\
     Mg\,{\sc{i}} &2&12&[Ne]$\mathrm{3s3p}$  & $\mathrm{^{3}P^{\circ}}$&  2512.26& & 21877.30  &9 & OP\\
     Mg\,{\sc{i}} &3&  12&[Ne]$\mathrm{3s3p}$  &  $\mathrm{^{1}P^{\circ}}$& 3755.50& & 35051.26& 3   & OP\\ 
     Mg\,{\sc{ii}} &1&  12&[Ne]$\mathrm{3s}$ &  $\mathrm{^{2}S}$&  & & 0.00 &2 & -\\    

		Al\,{\sc{i}} &1& 13&[Ne]$\mathrm{3s^2 3p}$&$\mathrm{^{2}P}^{\circ}$& 2077.19& 48278.20& 74.71 & 6 & OP\\ 
     Al\,{\sc{ii}} &1&  13&[Ne]$\mathrm{3s^2}$ & $\mathrm{^{1}S}$&  && 0.00 & 1  & - \\    
		Si\,{\sc{i}} &1& 14&[Ne]$\mathrm{3s^2 3p^2}$		&$\mathrm{^{3}P}$&1527.92& 65747.76&   149.68 & 9 & OP\\ 
		Si\,{\sc{i}} &2 & 14&[Ne]$\mathrm{3s^2 3p^2}$   &  $\mathrm{^{1}D}$& 1686.36&  &  6298.85 & 5&  OP\\  
		Si\,{\sc{i}} &3 & 14&[Ne]$\mathrm{3s^2 3p^2}$   &  $\mathrm{^{1}S}$ & 1991.88&  &  15394.37 & 1 & OP\\  
		Si\,{\sc{i}} &4 &14 &[Ne]$\mathrm{3s 3p^3}$     &  $\mathrm{^{5}S^{\circ}}$ & *\,1328.47 & & 33326.04 & 5  & OP\\      
		Si\,{\sc{i}} &5 & 14&[Ne]$\mathrm{3s^2 3p4s}$   &   $\mathrm{^{3}P^{\circ}}$&   3875.02& & 39799.50  & 9 & OP\\
		Si\,{\sc{i}} &6 & 14&[Ne]$\mathrm{3s^2 3p4s}$   &   $\mathrm{^{1}P^{\circ}}$&    4062.78&& 40991.88 & 3& OP\\ 
		Si\,{\sc{ii}} &1&  14&[Ne]$\mathrm{3s^2(^1S)3p}$ &   $\mathrm{^{2}P^{\circ}}$&  &&   0.00& 6 &  -\\
S\,{\sc{i}} &1& 16&[Ne]$\mathrm{3s^2 3p^4}$& $\mathrm{^{3}P}$& 1199.85&  83559.10& 107.74 &9 & OP\\
S\,{\sc{i}} &2&16&[Ne]$\mathrm{3s^2 3p^4}$&$\mathrm{^{1}D}$  & 1347.48&		 & 9238.61&5& OP\\
S\,{\sc{i}} &3&16&[Ne]$\mathrm{3s^2 3p^4}$  & $\mathrm{^{1}S}$&1632.08& 	 & 22179.95&1& OP\\   
S\,{\sc{ii}} &1& 16&[Ne]$\mathrm{3s^2 3p^3}$  & $\mathrm{^{4}S^{\circ}}$&  & & 0.00 &4 & - \\    
P\,{\sc{i}} &1&15& [Ne]$\mathrm{3s^2 3p^3}$ & $\mathrm{^{4}S^{\circ}}$ & 1182.30 &  84580.83& 0.0 & 4 &  Hyd\\
P\,{\sc{ii}}&1&15& [Ne]$\mathrm{3s^2 3p^2}$ & $\mathrm{^{3}P}$         &  & & 0.0 		    & 9 & - \\
Cl\,{\sc{i}} &1&17& [Ne]$\mathrm{3s^2 3p^5}$  & $\mathrm{^2P^{\circ}}$ &961.51  & 104591.0 & 0.0& 6 & Hyd\\
Cl\,{\sc{ii}}&1&17& [Ne]$\mathrm{3s^2 3p^4}$  & $\mathrm{^3P}$ &  &  & 0.0& 9 & - \\
Ar\,{\sc{i}}  &1&18& [Ne]$\mathrm{3s^2 3p^6}$ & $\mathrm{^1 S}$ &786.72 & 127109.80 & 0.0 & 1 &Hyd \\
Ar\,{\sc{ii}} &1&18& [Ne]$\mathrm{3s^2 3p^5}$ & $\mathrm{^2P^{\circ}}$ &  &    & 0.0 & 6 & - \\
K\,{\sc{i}} &1& 19&[Ar]$\mathrm{4s}$ &$\mathrm{^{2}S}$	&  2855.56&  35009.81&    0.00 &2 &  Hyd\\
K\,{\sc{i}} &2& 19&[Ar]$\mathrm{4p}$ &$\mathrm{^{2}P^{\circ}}$& 4547.06&  & 13023.64&6 & Hyd\\
K\,{\sc{i}} &3& 19&[Ar]$\mathrm{5s}$ &$\mathrm{^{2}S}$& 7149.41& & 21026.55& 2   & Hyd\\ 
K\,{\sc{i}} &4& 19&[Ar]$\mathrm{3d}$ &$\mathrm{^{2}D}$&   7419.51& & 21535.60&10   & Hyd\\ 
K\,{\sc{i}} &5& 19&[Ar]$\mathrm{5p}$ &$\mathrm{^{2}P^{\circ}}$&   9709.85& & 24713.90& 6   & Hyd\\ 
K\,{\sc{i}} &6& 19&[Ar]$\mathrm{4d}$ &$\mathrm{^{2}D}$&  13132.87& & 27397.50&10   & Hyd\\ 
K\,{\sc{i}} &7& 19&[Ar]$\mathrm{6s}$ &$\mathrm{^{2}S}$&13225.32& & 27450.71& 2   & Hyd\\ 
K\,{\sc{i}} &8& 19&[Ar]$\mathrm{4f}$ &$\mathrm{^{2}F^{\circ}}$&  14526.59& & 28127.85&14   & Hyd\\ 
K\,{\sc{i}} &9& 19&[Ar]$\mathrm{6p}$ &$\mathrm{^{2}P^{\circ}}$&  16648.27& & 29004.90& 6   & Hyd\\ 
K\,{\sc{i}} &10& 19&[Ar]$\mathrm{5d}$ &$\mathrm{^{2}D}$& 20721.96& & 30185.40&10   & Hyd\\ 
K\,{\sc{ii}} &1& 19&[Ar] 	      &  $\mathrm{^{1}S}$&  & & 0.00 &1 & - \\    
Ca\,{\sc{i}} &1& 20&[Ar]$\mathrm{4s^2}$& $\mathrm{^{1}S}$	&  2027.60&  49305.95&    0.00 &1& OP \\
Ca\,{\sc{i}} &2&20&[Ar]$\mathrm{4s4p}$  & $\mathrm{^{3}P^{\circ}}$&  2936.66&		&15263.09&9& OP\\
Ca\,{\sc{i}} &3& 20&[Ar]$\mathrm{3d4s}$&$\mathrm{^{1}D^{\circ}}$& 3641.13&   &    21849.63& 5  & OP\\ 
Ca\,{\sc{ii}} &1& 20&[Ar]$\mathrm{4s}$ &  $\mathrm{^{2}S}$&  & & 0.00 &2& - \\    
Sc\,{\sc{i}}  &1&21& [Ar]$\mathrm{3d4s^2}$  & $\mathrm{^{2}D}$ & 1896.81 & 52922.0& 0.0  & 10  & Hyd\\
Sc\,{\sc{ii}} &1&21& [Ar]$\mathrm{3d4s}$  & $\mathrm{^{3}D}$ &  & & 0.0 & 15 & -\\
Ti\,{\sc{i}}  &1&22& [Ar]$\mathrm{3d^2 4s^2}$& $\mathrm{^{3}F}$ & 1831.05 & 55010.0& 0.0 & 21   &   Hyd\\
Ti\,{\sc{ii}} &1&22& [Ar]$\mathrm{3d^2 4s}$  & $\mathrm{^{4}F}$ &  &        & 0.0 & 28   &   -\\
V\,{\sc{i}}  &1&23& [Ar]$\mathrm{3d^3 4s^2}$  & $\mathrm{^{4}F}$ & 1861.26 & 54360.0& 316.484    &30  & Hyd\\
V\,{\sc{ii}} &1&23& [Ar]$\mathrm{3d^4}$  & $\mathrm{^{5}D}$ &  & & 0.0 & 25 & -\\
      \noalign{\smallskip}
      \hline
  \end{tabular*}
  }}
\end{center}
\caption{Continued.} 
\end{table*}
\begin{table*}[!hh]  
\begin{center}
\addtocounter{table}{-1}
{\normalsize{
    \begin{tabular*}{.85\textwidth}{lrrllrrrrc}
      \hline
      \hline
      {{Ion}}  & 
      {{Level}}  &
      {{AN}}  & 
      {{Configuration}} &
      {{Term}} & 
      {{$\lambda_{\mathrm{ion}}$ (\AA)}}&  
      {{$E_{\mathrm{ion}}$} ($\mathrm{cm^{-1}}$)} &
      {{$\bar{E}$} ($\mathrm{cm^{-1}}$)}&
      {{$\bar{g}$}}&
      {{cross}}\\  
      \hline
\noalign{\smallskip}
Cr\,{\sc{i}}  &1&24& [Ar]$\mathrm{3d^5 4s}$  & $\mathrm{^{7}S}$ & 1832.32 &54575.6  &  0.0   & 7& Hyd\\
Cr\,{\sc{ii}} &1&24& [Ar]$\mathrm{3d^5}$  & $\mathrm{^{6}S}$ &  & & 0.0 &6 & -\\
Mn\,{\sc{i}}  &1&25& [Ar]$\mathrm{3d^5 4s^2}$  & $\mathrm{^{6}S}$ &1667.80  & 59959.40&  0.0   &6 & Hyd\\
Mn\,{\sc{ii}} &1&25& [Ar]$\mathrm{3d^5 4s}$  & $\mathrm{^{7}S}$ &  & & 0.0 &7 & -\\
Fe\,{\sc{i}} &1&26&[Ar]$\mathrm{3d^6 4s^2}$&$\mathrm{^5D}$ &  1589.04&  63737.00&    402.96 &25& OP \\
    Fe\,{\sc{i}} &2&26&[Ar]$\mathrm{3d^7 (^4F)4s}$ &  $\mathrm{^5F}$ &  1789.67&   &   7457.82 &35 &OP\\
    Fe\,{\sc{i}} &3&26&[Ar]$\mathrm{3d^7(^4F)4s}$ &  $\mathrm{^3F}$ &  *\,1873.48  & &   12408.10 & 21 &OP\\
    Fe\,{\sc{i}} &4&26&[Ar]$\mathrm{3d^7(^4P)4s}$ &  $\mathrm{^{5}P}$ &  2193.69   & &    17761.40& 15 &OP\\
    Fe\,{\sc{i}} &5&26&[Ar]$\mathrm{3d^6 4s^2}$  &    $\mathrm{^{3}P}$ &   2266.85&  & 19232.20 &9& OP\\
     Fe\,{\sc{ii}} &1&  26&[Ar]$\mathrm{3d^6 (^5D)4s}$ & $\mathrm{^{6}D}$& & &  0.00 &30 & - \\
     Co\,{\sc{i}}  &1&27& [Ar]$\mathrm{3d^7 4s^2}$  & $\mathrm{^{4}F}$ & 1616.96 & 63430.0&792.80 & 28 & Hyd\\
     Co\,{\sc{ii}} &1&27& [Ar]$\mathrm{3d^8}$       & $\mathrm{^{3}F}$ &  & & 0.0 & 21 & -\\
     Ni\,{\sc{i}}  &1&28& [Ar]$\mathrm{3d^8 4s^2}$  & $\mathrm{^{3}F}$ & 1676.27 & 61600.0 & 971.80& 21 & Hyd\\
     Ni\,{\sc{ii}} &1&28& [Ar]$\mathrm{3d^9}$       & $\mathrm{^{2}D}$ &  & & 0.0 & 10 &-\\
     Cu\,{\sc{i}}  &1&29& [Ar]$\mathrm{3d^10 4s}$   & $\mathrm{^{2}S}$ & 1604.69 & 62317.44 & 0.0    &2& Hyd\\
     Cu\,{\sc{ii}} &1&29& [Ar]$\mathrm{3d^10}$      & $\mathrm{^{1}S}$ &  &          & 0.0 &1& - \\
     Zn\,{\sc{i}}  &1&30& [Ar]$\mathrm{3d^10 4s^2}$  & $\mathrm{^{1}S}$ & 1319.80 & 75769.33& 0.0   &1& Hyd\\
     Zn\,{\sc{ii}} &1&30& [Ar]$\mathrm{3d^10 4s}$    & $\mathrm{^{2}S}$ &  &          & 0.0        &2& - \\
      \noalign{\smallskip}
      \hline
  \end{tabular*}
  }}
\end{center}
\caption{Continued.} 
\end{table*}

\end{appendix}
\end{document}